\documentclass[a4paper,onecolumn,11pt,unpublished]{quantumarticle}
\pdfoutput=1
\usepackage[utf8]{inputenc}
\usepackage[english]{babel}
\usepackage[T1]{fontenc}
\usepackage{amsmath}
\usepackage[hidelinks]{hyperref}
\usepackage[usestackEOL]{stackengine}

\usepackage{pgfplots}
\usepgfplotslibrary{colormaps}

\usepackage{tikz}
\usetikzlibrary{shapes.geometric}
\usetikzlibrary{quantikz2}

\pgfkeys{/tikz/.cd,
  re/.store in=\re,
  re=1.1547
}

\pgfkeys{/tikz/.cd,
  rd/.store in=\rd,
  rd=1.414213562
}

\pgfkeys{/tikz/.cd,
  shiftcz/.store in=\shiftcz,
  shiftcz=0.75
}

\pgfkeys{/tikz/.cd,
  shiftczz/.store in=\shiftczz,
  shiftczz=-1
}

\definecolor{codered}{rgb}{0.9, 0.05, 0.05}
\definecolor{codegreen}{rgb}{0.2, 0.6, 0.2}
\definecolor{codeblue}{rgb}{0.15, 0.15, 0.7}
\definecolor{roughcolor}{RGB}{255, 187, 0}
\definecolor{smoothcolor}{RGB}{0, 187, 255}

\usepackage{lipsum}
\usepackage{subcaption}
\usepackage{stmaryrd}

\newcommand{\topwrite}[4][3pt]{
  \gategroup[#2,steps=#3,style={inner
xsep=0pt, inner ysep=#1, white!1},background,label style={label
position=above,anchor=north}]{#4}}
\newcommand{\topwriterough}[4][3pt]{
  \gategroup[#2,steps=#3,style={inner
xsep=0pt, inner ysep=#1, roughcolor!50},background,label style={label
position=above,anchor=north}]{#4}}
\newcommand{\meterX}{\meter[label style={yshift=-0.7cm}]{\text{\tiny $X$}}}
\newcommand{\meterZ}{\meter[label style={yshift=-0.7cm}]{\text{\tiny $Z$}}}

\newcommand{\fig}[1]{(see Fig.\,\ref{fig:#1})}
\newcommand{\secc}[1]{Sec.\,\ref{sec:#1}}
\newcommand{\XX}{$X$-controlled NOT}

\begin{document}

\title{Fault-Tolerant Constant-Depth Clifford Gates on Toric Codes}

\author{Alexandre Guernut}
\affiliation{Universit\'e de Lorraine, CNRS, Inria, LORIA, F-54000 Nancy, France}

\author{Christophe Vuillot}
\affiliation{Universit\'e de Lorraine, CNRS, Inria, LORIA, F-54000 Nancy, France}

\maketitle

\begin{abstract}
	We propose and simulate the performance of a set of fault-tolerant and constant-depth logical gates on 2D toric codes.
	This set combines fold-transversal gates, Dehn twists and single-shot logical Pauli measurements and generates the full Clifford group.
\end{abstract}

\section{Introduction}

Building large scale and reliable quantum computers requires the use of quantum error correction and fault-tolerant quantum gates.
Such techniques introduce a large resource overhead as demonstrated by resource estimation for the leading candidate, the surface code using lattice surgery \cite{kitaevFaulttolerantQuantumComputation2003, dennisTopologicalQuantumMemory2002, horsmanSurfaceCodeQuantum2012, gidneyHowFactor20482021}.
Quantum error correcting codes encoding several logical qubits per code-block enable a more efficient use of resources \cite{gottesmanFaulttolerantQuantumComputation2014}.
One of the difficulty introduced by such codes is the implementation of logical gates on targeted logical qubits, as well as their parallelization.
Some generic methods have been devised such as state preparation and gate teleportation \cite{knillScalableQuantumComputing2005, gottesmanDemonstratingViabilityUniversal1999} as well as generic ways of performing code surgery \cite{cohenLowoverheadFaulttolerantQuantum2022}.
Standard surgery methods come with a time overhead as they rely on repeating measurements a number of time scaling with the distance of the code.
Recently there has been a renewed interest in finding ways around this time overhead by the prospect of more flexible connectivity between qubits within architectures with movable qubits, see e.g. \cite{bluvsteinLogicalQuantumProcessor2024}.
The use of transversal gates \cite{zhouAlgorithmicFaultTolerance2024} as well as techniques for single-shot measurements \cite{vuillot_fault-tolerant_2020, huangHomomorphicLogicalMeasurements2023, hillmannSingleshotMeasurementbasedQuantum2024} have been proposed.

In this work we combine transversal and fold-transversal gates \cite{moussaTransversalCliffordGates2016, breuckmannFoldTransversalCliffordGates2024}, with single-shot logical measurements \cite{huangHomomorphicLogicalMeasurements2023} and Dehn-twists \cite{breuckmannHyperbolicSemihyperbolicSurface2017}.
Applying these techniques on 2D toric codes \cite{kitaevFaulttolerantQuantumComputation2003} we obtain a set of constant-depth logical gates generating the full Clifford group.
In Section~\ref{sec:design} we describe our set of logical gates, then in Section~\ref{sec:simulation} we simulate and estimate the performance of these schemes.

\section{Constant Depth Logical Gates on Toric Codes}
\label{sec:design}

In this section we present a set of constant-depth and fault-tolerant logical gates for toric codes which generate the full Clifford group.
We consider a setting where the connectivity is not a constraint and we allow rearranging the qubits for free when needed.

\begin{figure*}
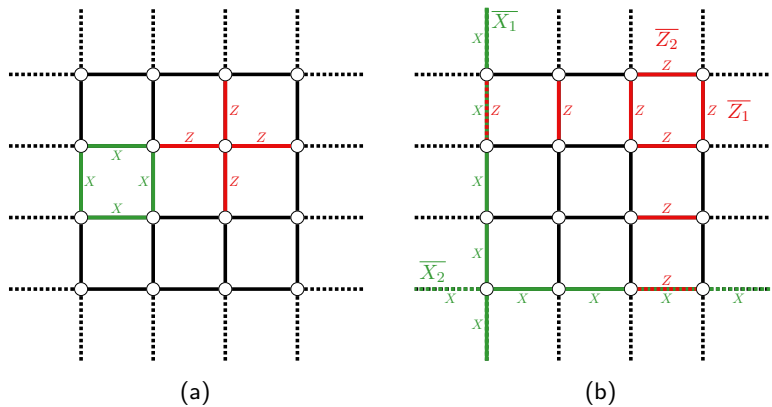

  \centering
  \begin{subfigure}{0.33\linewidth}
    \resizebox{\linewidth}{!}{%
  \input{./figs/baretoric.tikz}%

    }
    \subcaption{}
    \label{fig:fig:TC1}
  \end{subfigure}\quad
  \begin{subfigure}{0.33\linewidth}
    \resizebox{\linewidth}{!}{%
  \input{./figs/baretoric2.tikz}%

    }
    \subcaption{}
    \label{fig:fig:TC2}
  \end{subfigure}
  \caption{Toric code with \texorpdfstring{$d=4$}~. Dotted edges on opposite sides of the square are identified. (\subref{fig:fig:TC1}) \(X\) stabilizers are products of operators acting as \(X\) on the four qubits of a face. \(Z\) stabilizers are products of operators acting as \(Z\) on the the four qubits linked to a vertice. (\subref{fig:fig:TC2}) Four independent Pauli operators commutes with every stabilizers. They form the two pairs of anticommuting logical Pauli operators of the code: \((\overline{X_1}, \overline{Z_1})\) corresponding to logical qubit 1, and \((\overline{X_2}, \overline{Z_2})\) corresponding to logical qubit 2.}
  \label{fig:TC}
\end{figure*}

We work with the standard toric code, parameterized by its code distance $d$, defined on a square lattice embedded on a torus, with parameters $\llbracket 2d^2, 2, d \rrbracket$ and denote it as $\mathcal{C}$.
For the stabilizers of $\mathcal{C}$ we pick the following convention: $X$-stabilizers act on the four qubits adjacent to each face and $Z$-stabilizers act on the four qubits adjacent to each node, see~Figure~\ref{fig:fig:TC1}.
In this convention, $X$ logical operator run along non-trivial cycles and $Z$ logical operators run along non-trivial co-cycles, see~Figure~\ref{fig:fig:TC2}.

\subsection{Transversal Two-Qubit Gates Across Two Blocks of Code}
\label{sec:sec:trans2qb}

\begin{figure*}
  \centering
  \begin{subfigure}{0.45\linewidth}
    \resizebox{\linewidth}{!}{%
  \input{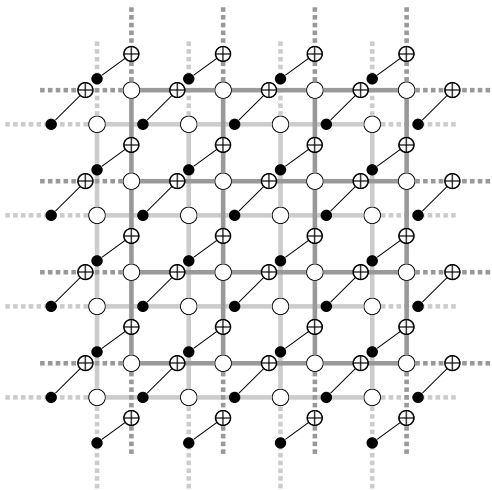}%

    }
    \subcaption{Transversal CNOT gate.}
    \label{fig:fig:transCNOT}
  \end{subfigure}
  \hfill
  \begin{subfigure}{0.45\linewidth}
    \resizebox{\linewidth}{!}{%
  \input{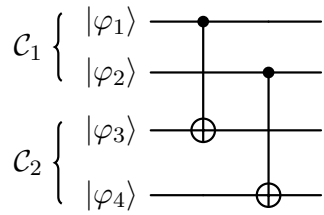}%

    }
    \subcaption{Transversal C$Z$ gate.}
    \label{fig:fig:transCZ}
  \end{subfigure}

  \begin{subfigure}{0.45\linewidth}
    \hspace{0.16\linewidth}
    \resizebox{0.66\linewidth}{!}{%
  \input{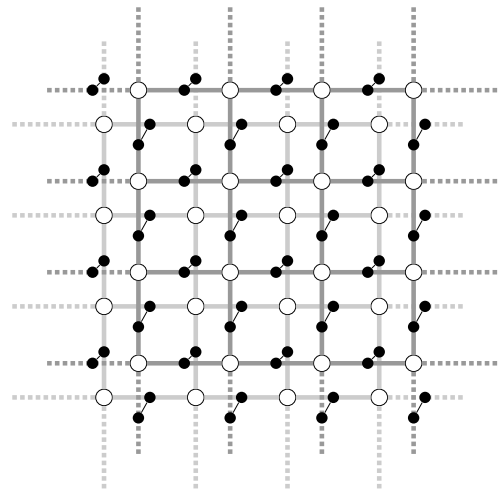}%

    }
    \subcaption{Logical circuit of the transversal CNOT.}
  \end{subfigure}
  \hfill
  \begin{subfigure}{0.45\linewidth}
    \hspace{0.16\linewidth}
    \resizebox{0.66\linewidth}{!}{%
  \input{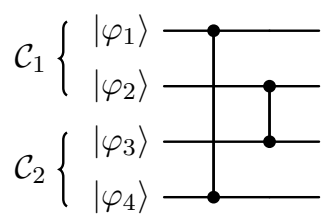}%

    }
    \subcaption{Logical circuit of the transversal C$Z$.}
  \end{subfigure}
  \caption{Transversal CNOT and C$Z$ gates on toric codes.}
  \label{fig:trans}
\end{figure*}

A toric code can be entangled with a surface code \cite{bravyiQuantumCodesLattice1998} or with another toric code using some transversal two-qubit gates, if the underlying lattices are similar. The transversal CNOT gate between two toric codes is probably the most well-known example, but others configurations are possible. The stabilizer group must be preserved by the procedure. When using only transversal two-qubit gates, this means stabilizer operators of either code must only be mapped to products of stabilizer operators of these codes.

The logical information is not threatened in the process, despite physical error being potentially locally spread from one code to the other. Indeed, both logical operators and physical errors that flip them spread at the same time and in opposite direction, in such a way that if $d$ errors were needed to flip the logical information without the transversal gates, $d$ errors are still needed despite them. An operator of weight \(d\) might spread across, becoming an operator of weight \(2d\) \textit{i.e.} requiring twice as much errors to be flipped. Conversely, some product operator of weight \(2d\) will be shortened to be of weight \(d\) \textit{i.e.} still requiring at least \(d\) errors to be flipped. The overall distance of \(d\) is not decreased, although some high-weight operator can see there weight decrease towards \(d\).

Hence, the transversal application of two-qubit gates is a fault-tolerant operation, provided that stabilizer operators are preserved. Intuitively, $X$-stabilizers spread through controls (•) and $Z$-stabilizers spread through targets ($\oplus$). These open the path for other transversal two-qubit operations that goes beyond the transversal CNOT gate between two toric codes. Such an operation can be done in one time step, making it a constant time operation with respect to the distance of the code.

\subsubsection{Transversal C$Z$ gate and alike}

The toric code being a CSS code, transversal CNOTs across two toric codes will perform a logical transversal CNOT on the two pairs of logical qubits, \fig{fig:transCNOT}.
This can be checked directly as $X$-stabilizers of the controlled code are mapped to their product with their $X$ counterpart in the targeted code while $Z$-stabilizers of the targeted code are mapped to their product with their $Z$ counterpart in the controlled code. Other stabilizers (respectively $Z$-stabilizers of the controlled code and $X$-stabilizers of the targeted code) remain identical and hence the stabilizer group is preserved.
Logical operators are also mixed up in the same fashion: some of them have grown in the other code.

A transversal C$Z$ gate between two toric codes can also be designed \fig{fig:transCZ}. The C$Z$ are controlled such that any $X$-stabilizer of weight 4 has a unique corresponding $Z$-stabilizer on the other toric code as target. Hence, stabilizers of both codes are mapped to product of stabilizers. In the same way, $X$ logical Pauli operators of one code are mapped to their product by the $Z$ logical operator along the same direction on the tori. A very similar gate can be obtained by using \XX~gates instead of C$Z$ gates.

It is worth noting that the two logical two-qubit gates arising from either transversal implementation have the same support and that this support is different from the one found for the transversal CNOT gate. This is not surprising, as these two gates can be obtained by applying a transversal Hadamard gate (see \secc{sec:transH}) on one code or the other, swapping the two logical qubits on one side.

\subsubsection{Toric code entanglement with cylindrical patches}
\label{sec:sec:sec:patch}

\begin{figure*}
  \centering
  \resizebox{0.4\textwidth}{!}{%
    \includegraphics{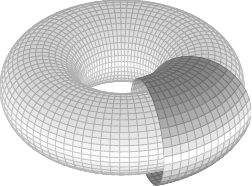}
  }
  \caption{Schematic spatial disposition of a toric code and a surface code cylindrical patch.}
  \label{fig:manchon}
\end{figure*}

Interacting with only a subset of the logical qubits of a block code can be challenging. One logical qubit of a toric code can be entangled with the unique logical qubit of a cylindrical patch via the transversal application of two-qubit gates between the two codes \fig{manchon}. Such a patch can easily be obtained by measuring a subset of qubits of a toric code (along a logical Pauli operator), thus opening a boundary. Alternatively, this cylinder can be obtained from a regular patch of surface code by fusing the two  \(X\) (resp. \(Z\)) boundaries together, such that the $Z$ logical Pauli operator (resp. $X$ logical Pauli operator) of the code lie on its non trivial loop. While this preparation requires several time steps, the time cost can be amortized by preparing the needed patches ahead of time.

\begin{figure*}
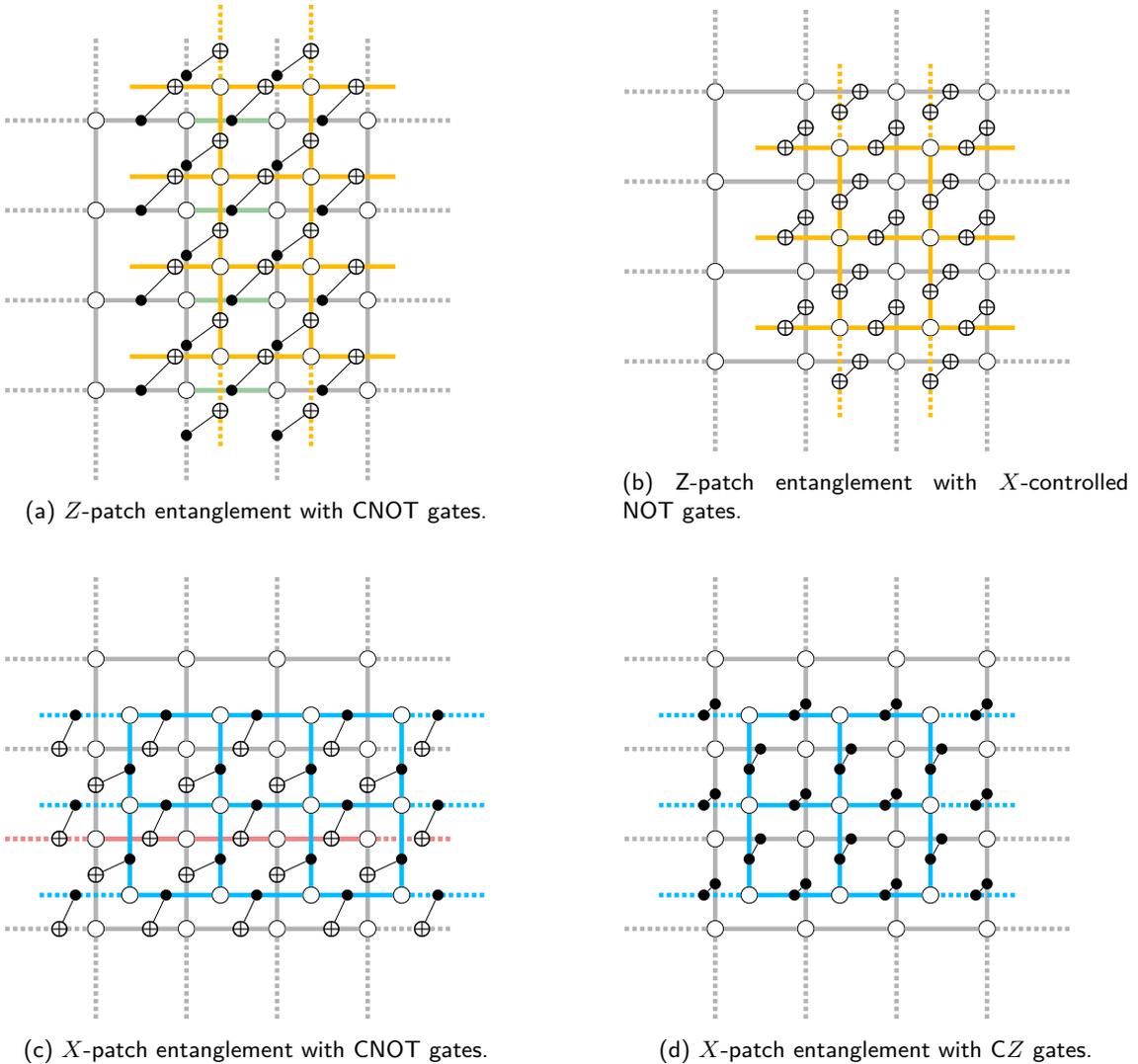

  \centering
  \begin{subfigure}{0.45\linewidth}
    \resizebox{\linewidth}{!}{%
  \input{./figs/rough.tikz}%

    }
    \subcaption{\(Z\)-patch entanglement with CNOT gates.}
    \label{fig:fig:rough}
  \end{subfigure}
  \hfill
  \begin{subfigure}{0.45\linewidth}
    \resizebox{\linewidth}{!}{%
  \input{./figs/rough2.tikz}%

    }
    \subcaption{Z-patch entanglement with \XX~gates.}
    \label{fig:fig:rough2}
  \end{subfigure}

  \begin{subfigure}{0.45\linewidth}
    \resizebox{\linewidth}{!}{%
  \input{./figs/smooth.tikz}%

    }
    \subcaption{\(X\)-patch entanglement with CNOT gates.}
    \label{fig:fig:smooth}
  \end{subfigure}
  \hfill
  \begin{subfigure}{0.45\linewidth}
    \resizebox{\linewidth}{!}{%
  \input{./figs/smooth2.tikz}%

    }
    \subcaption{\(X\)-patch entanglement with C$Z$ gates.}
    \label{fig:fig:smooth2}
  \end{subfigure}
  \caption{Possible entanglement between the first qubit of a toric code and cylindrical patches. (\subref{fig:fig:rough}) The logical {\color{codegreen!50}green} $Z$ Pauli operators would be measured should the \(Z\)-patch be measured. (\subref{fig:fig:smooth}) The logical {\color{codered!50}red} $X$ Pauli operator would be measured should the \(X\)-patch be measured.}
  \label{fig:roughsmooth}
\end{figure*}

  The boundaries of the patch must be chosen in accordance with the gates to be applied in order to preserve the set of stabilizers. Namely, a patch with \(Z\) boundaries cannot support any control, as the truncated $X$-stabilizer operators on the boundaries will be mapped to non-stabilizer Pauli operators (their counterpart has a different weight) while a patch with \(X\) boundaries cannot support any target, as the truncated $Z$-stabilizer operators on the boundaries will be mapped to non-stabilizer Pauli operators on the other side. The Figure~\ref{fig:roughsmooth} shows the four allowed procedures involving cylindrical patches. When C$Z$ (or \XX) gates are used, a small shift as to be made between the two lattices, just like in the previous case \fig{fig:transCZ}.

The patch must be large enough to hinder neither the $X$ distance nor the $Z$ distance of the qubit (or at least not too much), preferably a $d \times (d-1)$ cylindrical patch for a $d \times d$ toric code. The \(Z\)-patch (resp. \(X\)-patch) is aligned along one representative of a $Z$ (resp. $X$) logical Pauli operators of the logical qubit of interest.

It is important to note that the other qubits of the block code will not be entangled with the patch, as the latter only support one logical qubit. While it is true that some representatives of the logical Pauli operators of the other logical qubit might then pass through the patch, measuring it would not provide any information about the state of these logical qubits.

These operations only consist of a single layer of concurrent transversal two-qubit gates, hence they all are constant time operations with respect to the code distance.

\subsection{Logical qubit displacement}
\label{sec:sec:logqbtp}

Logical qubit routing is likely to be needed when compiling a quantum circuit into an hardware-compatible circuit. The transversal SWAP gate can help a bit with moving around block of qubits, but cannot move individual qubits. In order to do so, we propose to use the one-bit teleportation techniques introduced in \cite{zhouMethodologyQuantumLogic2000} alongside the one-qubit entanglement procedures from the previous section.

\begin{figure*}
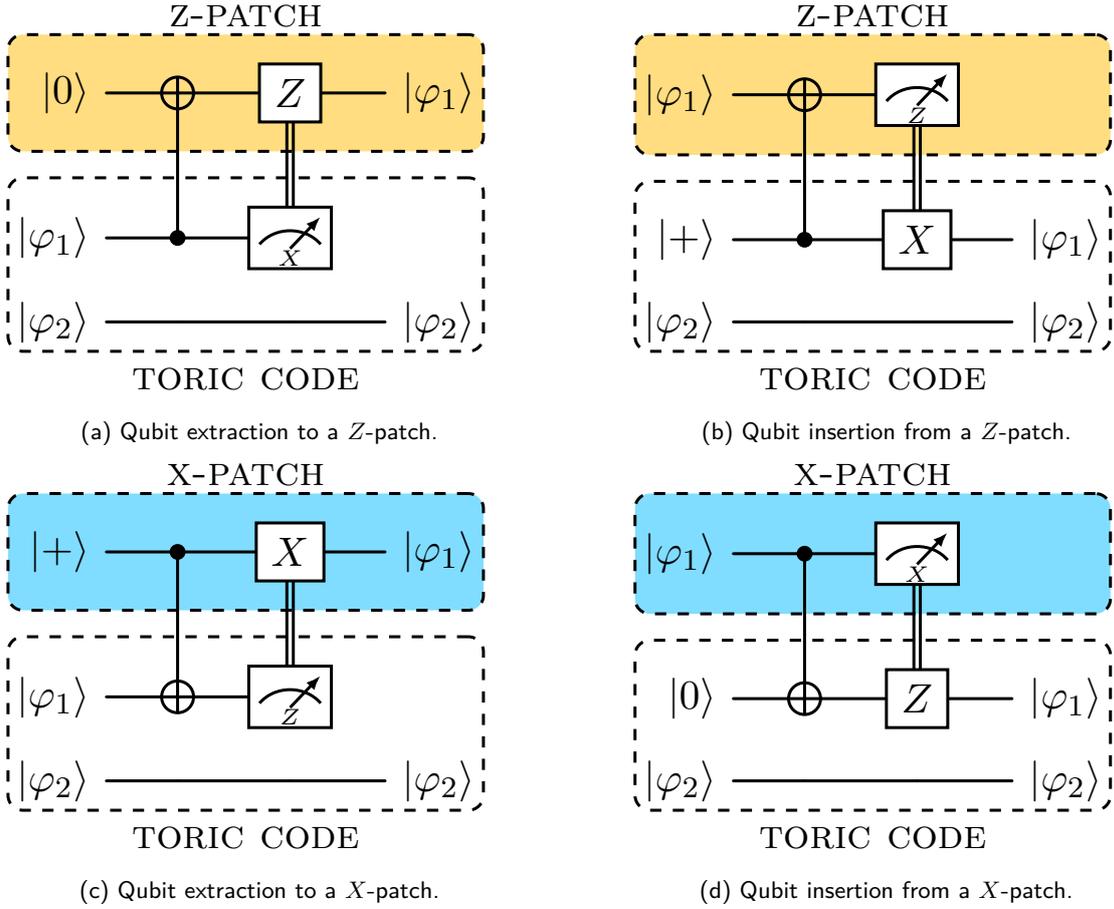

  \centering
  \begin{subfigure}{0.45\linewidth}
    \resizebox{\linewidth}{!}{%
  \input{./figs/tp1.tikz}%

    }
    \subcaption{Qubit extraction to a \(Z\)-patch.}
    \label{fig:fig:tp1}
  \end{subfigure}
  \hfill
  \begin{subfigure}{0.45\linewidth}
    \resizebox{\linewidth}{!}{%
  \input{./figs/tp2.tikz}%

    }
    \subcaption{Qubit insertion from a \(Z\)-patch.}
    \label{fig:fig:tp2}
  \end{subfigure}

  \begin{subfigure}{0.45\linewidth}
    \resizebox{\linewidth}{!}{%
  \input{./figs/tp3.tikz}%

    }
    \subcaption{Qubit extraction to a \(X\)-patch.}
    \label{fig:fig:tp3}
  \end{subfigure}
  \hfill
  \begin{subfigure}{0.45\linewidth}
    \resizebox{\linewidth}{!}{%
  \input{./figs/tp4.tikz}%

    }
    \subcaption{Qubit insertion from a \(X\)-patch.}
    \label{fig:fig:tp4}
  \end{subfigure}
  \caption{Single qubit displacement from and toward a toric code.}
  \label{fig:qbtp}
\end{figure*}

Qubit extraction require the ability to measure one specific qubit from a block code. How to do this will be shown in \secc{sec:partmeas} and requires a constant time procedure. Overall, qubit extraction is a constant depth and therefore constant time procedure with respect to the code distance. A partial reset of the target location is also needed, which can be done again with a single-qubit measurement, followed by a Pauli correction. Qubit insertion is also a constant time procedure, requiring only a CNOT round and a classically-controlled Pauli correction.

\subsection{Partial measurement}
\label{sec:sec:partmeas}

One disadvantage of block codes is the measurement of the logical state. Logical qubits are not easily measured individually, as directly measuring the physical qubits ends up collapsing the whole logical state.

\begin{figure*}
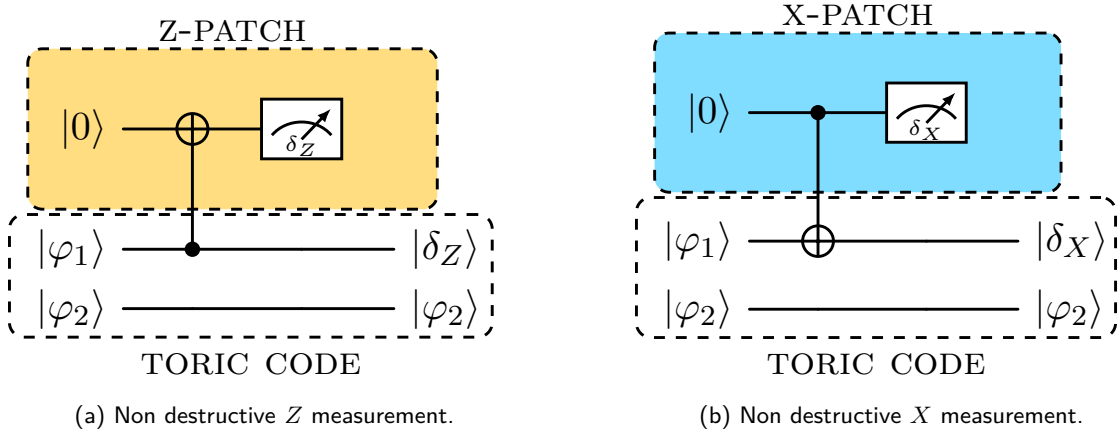

  \centering
  \begin{subfigure}{0.45\linewidth}
    \resizebox{\linewidth}{!}{%
  \input{./figs/OQM.tikz}%

    }
    \subcaption{Non destructive $Z$ measurement.}
    \label{fig:fig:measZ}
  \end{subfigure}
  \hfill
  \begin{subfigure}{0.45\linewidth}
    \resizebox{\linewidth}{!}{%
  \input{./figs/OQMX.tikz}%

    }
    \subcaption{Non destructive $X$ measurement.}
    \label{fig:fig:measX}
  \end{subfigure}
  \caption{Non destructive measurement used in the partial measurement protocol.}
  \label{fig:meas}
\end{figure*}

To measure logical qubits individually, we propose the use of an auxiliary patch of surface code shaped like a cylinder \fig{manchon}, similarly to what is suggested in \cite{huangHomomorphicLogicalMeasurements2023}. This patch is chosen and oriented to correspond to the logical Pauli operator of interest. It is then entangled with the block code, through means explained in \secc{sec:sec:patch}. As a consequence, the logical qubit to be measured and the one stored in the cylinder are entangled. The physical qubits of the cylinder are then measured in the adequate basis \fig{meas}, collapsing the logical qubit of the torus to a known -- now measured -- state.

The section of the cylinder must be the same as the one of the toric code, along the direction to be measured such that the transversal CNOT gates application make sense (see \secc{sec:sec:patch}). The height of the cylinder cannot be greater than the section of the torus in the other direction. In the absence of errors, it could be 1, making the cylinder a repetition code. The greater the height of the cylinder is, the better one will be at catching errors that would otherwise flip the measurement outcome. The Figures \ref{fig:fig:rough} and \ref{fig:fig:smooth} show respectively the measurement of the first qubit of a toric code in the $Z$ and $X$ basis, using a \(Z\)-patch of height 3 and a \(X\)-patch of height 3.

We repeat here our remark from \secc{sec:logqbtp}: despite the representatives of some logical operators of the non-measured logical qubits of the torus having their support modified to include some physical qubits of the cylinder, one cannot learn anything from their logical state. Indeed, these elongated logical operators, when restricted solely to the cylinder does not form a logical operator of the latter: their measurement outcome is random, hence no information is gained from them when measuring the cylinder. See \cite{huangHomomorphicLogicalMeasurements2023} for a more precise discussion of distance. This paper also offers means to measure joint logical operators, which we won't need here.

Setting aside the initialization of the patch which could be amortized for a longer computation, this procedure can be realized in constant time with respect to the distance of the code.

\subsection{Alternative CNOT gates}
We explained before how performing transversal CNOT gates between all physical qubits of two code blocks implements logical CNOT gates in Section~\ref{sec:sec:trans2qb}. While this is all that is needed for surface codes encoding a single logical qubit, it is cumbersome to work with when using block codes since it performs all pairwise CNOT gates between the logical qubits of the two blocks at once. Some other techniques are necessary to break the symmetry of the transversal CNOT gates.

\subsubsection{Dehn twist}
\label{sec:sec:sec:DT}
\begin{figure*}
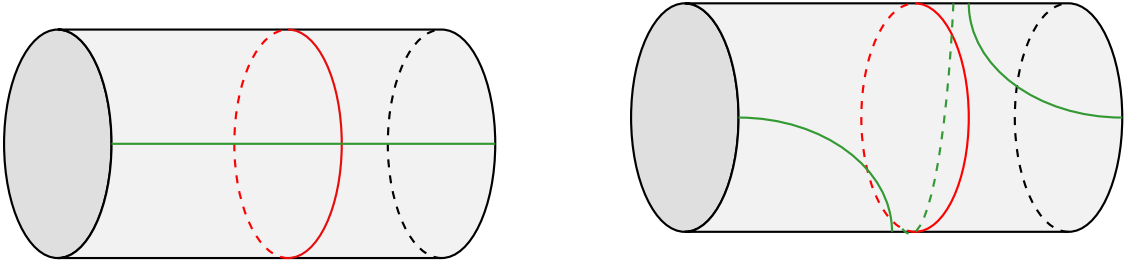

  \centering
  \begin{subfigure}{0.45\linewidth}
    \resizebox{\linewidth}{!}{%
  \input{./figs/dehncyl.tikz}%

    }
  \end{subfigure}
  \hfill
  \begin{subfigure}{0.45\linewidth}
    \resizebox{\linewidth}{!}{%
  \input{./figs/dehncyl2.tikz}%

    }
  \end{subfigure}
  \caption{A Dehn twist on a continuous cylinder. The {\color{codegreen}green} operator is twisted along the {\color{codered}red} guide.}
  \label{fig:cyl}
\end{figure*}

A Dehn twist guided along a closed curve of a surface is a self-automorphism of that surface whose curves intersecting the guide are twisted along that guide \cite{breuckmannHyperbolicSemihyperbolicSurface2017}. When a surface code is embedded on a surface, performing a Dehn twist along a closed loop bearing Pauli logical operators of the code will mix up these operators with the operators intersecting the guide support. By mapping closed loops of a surface code to product of closed loops of the same surface codes, a Dehn twist maps Pauli operators of a surface code to Pauli operators, performing a Clifford operation, often a set of CNOT gates \fig{cyl}.

\begin{figure*}
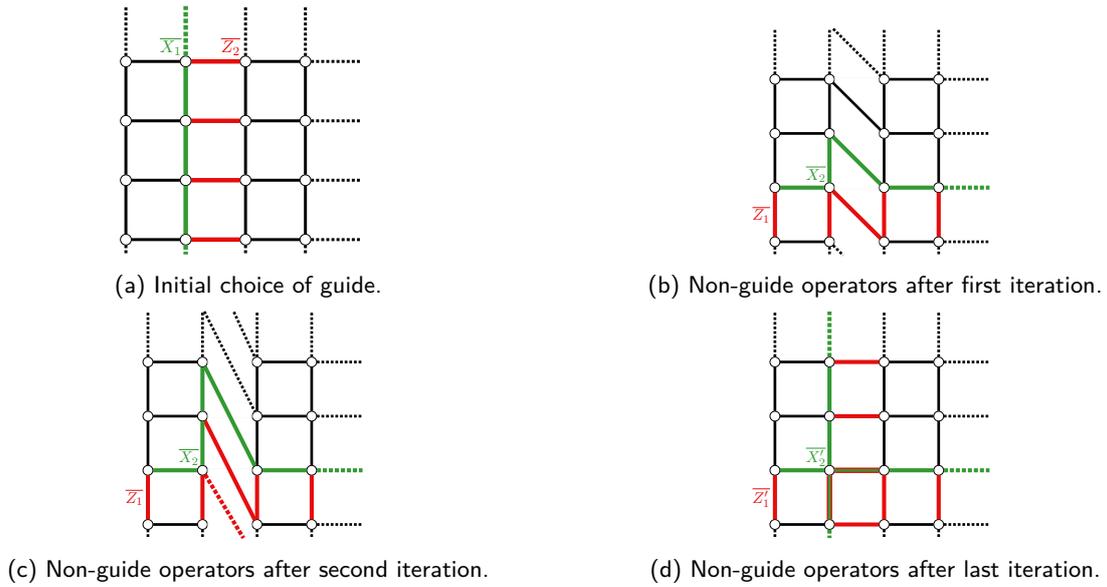

  \centering
  \begin{subfigure}{0.45\linewidth}
    \centering
    \resizebox{0.5\linewidth}{!}{%
  \input{./figs/dtstep1.tikz}%

    }
    \subcaption{Initial choice of guide.}
  \end{subfigure}
  \hfill
  \begin{subfigure}{0.45\linewidth}
    \centering
    \resizebox{0.5\linewidth}{!}{%
  \input{./figs/dtstep2.tikz}%

    }
    \subcaption{Non-guide operators after first iteration.}
  \end{subfigure}
  
  \begin{subfigure}{0.45\linewidth}
    \centering
    \resizebox{0.5\linewidth}{!}{%
  \input{./figs/dtstep3.tikz}%

    }
    \subcaption{Non-guide operators after second iteration.}
  \end{subfigure}
  \hfill
  \begin{subfigure}{0.45\linewidth}
    \centering
    \resizebox{0.5\linewidth}{!}{%
  \input{./figs/dtstep4.tikz}%

    }
    \subcaption{Non-guide operators after last iteration.}
  \end{subfigure}
  \caption{Different consecutive steps of a Dehn twist performed on a toric code.}
  \label{fig:DT}
\end{figure*}

While Dehn twists are usually defined in a continuous geometry setup, they can analogously be implemented on a particular discrete embedding of a finite distance surface code. One possible way to realize a Dehn twist has been shown in \cite{breuckmannHyperbolicSemihyperbolicSurface2017} and goes as follow:

\begin{enumerate}
\item Choose a representative $L_Z$ of a $Z$ logical Pauli operator as guide.
\item There exists a representative $L_X$ of a $X$ logical Pauli operator that runs parallel to the guide.
\item \label{cnotstep} Apply CNOT gates controlled on the guide and targeting the $X$ operator.
\item Update the stabilizers \fig{DT} and perform a correction step.
\item Repeat from step \ref{cnotstep} by shifting the controls in one direction along the guide until the original stabilizers are retrieved.
\end{enumerate}

The choice of $L_Z$ is arbitrary, though it is preferred to be as small as possible (ideally $\left| L_Z \right| = d$).
After a round of CNOT gates is applied, the set of stabilizer operators changes: the code space is slightly modified. The new set of stabilizers can still be interpreted as the faces and vertices of an embedding of the surface, defining a new surface code. The new embedding is very similar to the previous one, differing only by a twist along the guide direction. The logical operators of the previous code are mapped to those of the next one \fig{DT}, hence preserving the encoded state. The operators along the guide direction are untouched, operators crossing it are twisted just like the stabilizers. Each round of CNOT gates can both introduce new errors and slightly spread existing ones which should be periodically mitigated through error correction steps, typically after one or a few CNOT gate rounds. Each error correction step is performed using the newest support for the stabilizers, while the measurement outcome for each stabilizer is compared to the outcome previously measured for its antecedent.

After $\left| L_Z \right|$ CNOT gates rounds, the set of stabilizer operators is identical to the one at the start and so is the code space. However, every $X$ logical Pauli operator of the code crossing the guide has been multiplied by $L_X$ and every $Z$ logical Pauli operator of the code crossing the guide has been multiplied by $L_Z$ while logical Pauli operators along the guide remained unchanged. A circuit of logical CNOT gates within one block of surface code has thus been implemented. When the block encodes two logical qubits, it means a single CNOT gate has been applied between these two qubits.

The full Dehn twist procedure takes at least $d$ time steps for the CNOT gates and $d$ time steps per error correction round. Considering an error correction round with \(m\) measurement repetitions following every $\alpha$ CNOT rounds, the
overall minimal duration of the Dehn twist procedure is $d + m \left\lceil\frac{d}{\alpha}\right\rceil$ time steps. We expect this procedure to cut the code distance to \(\frac{d}{\alpha}\).

\subsubsection{Instantaneous Dehn twist}
\label{sec:sec:sec:IDT}

\begin{figure*}
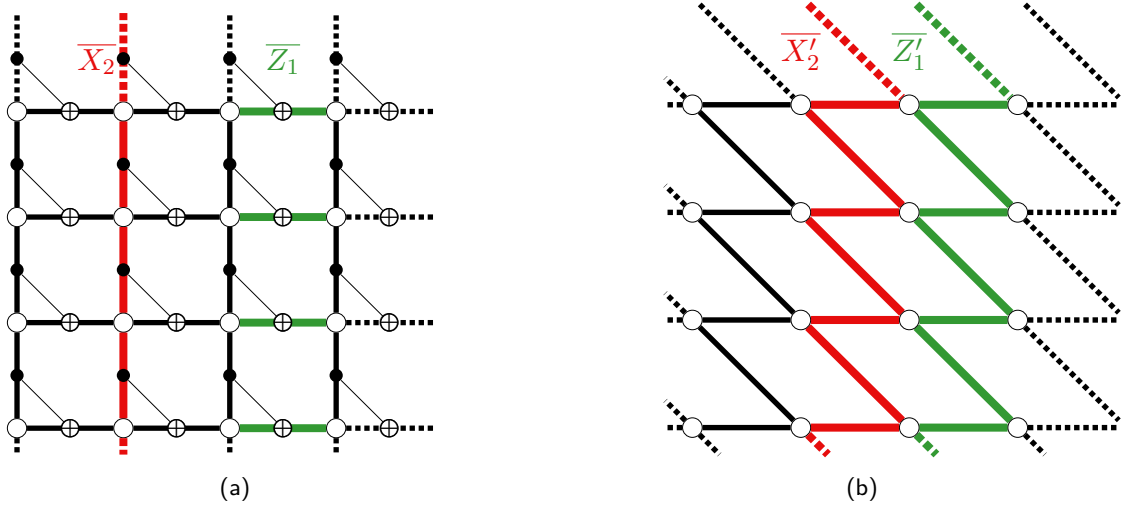

  \centering
  \begin{subfigure}{0.4\linewidth}
    \resizebox{\linewidth}{!}{%
  \input{./figs/IDT.tikz}%

    }
    \subcaption{}
    \label{fig:fig:IDT1}
  \end{subfigure}
  \hfill
  \begin{subfigure}{0.5\linewidth}
    \resizebox{\linewidth}{!}{%
  \input{./figs/IDT2.tikz}%

    }
     \subcaption{}
    \label{fig:fig:IDT2}
  \end{subfigure}
  \caption{An instantaneous Dehn twist protocol along horizontal guides (not depicted). \(X_2\) and \(Z_1\) are the non-guide logical operators, that become \(X_2^{\prime}=X_1X_2\) and \(Z_1^{\prime}.\)}
  \label{fig:IDT}
\end{figure*}

When considering the special case of a $d \times d$ square toric code, a simpler and shorter procedure can be used at the expense of a permanent change of stabilizers. Consider the application of CNOT gates controlled on every vertical qubit and targeting the qubit on the immediate bottom right \fig{fig:IDT1}. They can be interpreted as doing the first step of the Dehn twist procedure in parallel along $d$ representatives of the horizontal logical $Z$ Pauli operator. All the stabilizer operators are slightly distorted at once. The horizontal guide logical operators remain untouched, while one can check that vertical ones are mapped to the product of the two $Z$ (resp. $X$) logical operators of the distorted code \fig{fig:IDT2}.

It is also possible to act only on a portion $d \times d/k$ of a toric code (the rest of the code might be busy undergoing other procedures), while keeping some speed-up. With the larger direction as guide, by performing $k$ instantaneous Dehn twists as explained above, non-guide logical Pauli operators are fully twisted, one $k$-th of a turn at each step. Note that a particular $d \times d/k$ region of a toric code allows only for one guide direction.

When a whole $d \times d/k$ toric code is chosen instead, both directions are available. When choosing the smaller direction as guide, the instantaneous Dehn twist can be performed in one go: simply ignore any $d - d/k \times d/k$ region and apply CNOT gates as described above for the square case. From there we can readily describe the procedure for an arbitrary $a \times b$ toric code, with $a \geq b$: one apply one step of CNOT gates on a $b \times b$ cylinder for one direction or apply $k$ steps everywhere for the other direction, with $k$ chosen such that $a \leq kb < 2a$. Such a $k$ always exists.

It is important to note that the instantaneous procedure can be made involutive, if one alternates the side for the targeted qubits \textit{i.e.} choosing the immediate bottom \emph{left} qubits if the code underwent an odd number of instantaneous Dehn twists in the considered direction. This ensures the needed connectivity does not grow out of control as circuit depth increases, especially in the square toric code case. Moreover, transversal operations between a distorted toric code and a regular one are still possible, as the distortion is not local: stabilizers of both codes still have the same size and shape.

As the name indicates, the instantaneous Dehn twist is performed in constant time with respect to the distance of the code.

\subsection{Transversal Hadamard gate}
\label{sec:sec:transH}
\begin{figure*}
  \centering
  \begin{subfigure}{0.45\linewidth}
    \resizebox{\linewidth}{!}{%
  \input{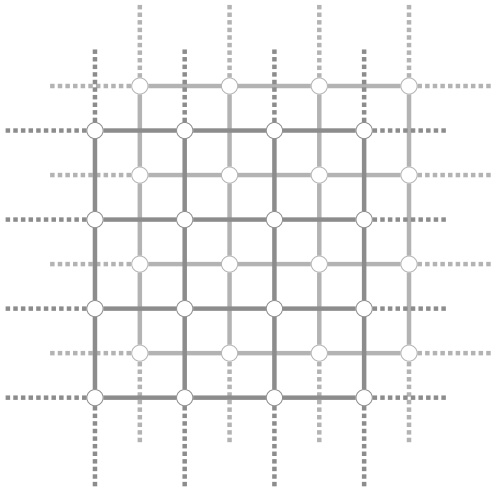}%

    }
    \subcaption{Stabilizers and qubit alignment {\color{gray!60}before} and {\color{gray!90}after} transversal Hadamard gates.}
    \label{fig:fig:transH}
  \end{subfigure}
  \hfill
  \begin{subfigure}{0.45\linewidth}
    \hspace{0.16\linewidth}
    \resizebox{0.66\linewidth}{!}{%
  \input{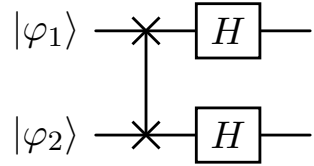}%

    }
    \subcaption{Logical circuit of the transversal Hadamard gate.}
    \label{fig:fig:transHlog}
  \end{subfigure}
  \caption{Transversal Hadamard gate on the toric code.}
  \label{fig:transH}
\end{figure*}

Applying an Hadamard gate on every physical qubits of a toric code is a well known transversal operation. Each $X$-stabilizer (resp. $Z$-stabilizer) is mapped to a $Z$ Pauli operator (resp. $X$ Pauli operator) on the same support, thus preserving locality. These Pauli operators can be laid on a toric square grid, which happen to be the dual of the original toric code layout \fig{fig:transH}. They can be thought as the stabilizers of a dual toric code laid upon the dual lattice. Original $X$ logical Pauli operators (resp. $Z$ Pauli operators) are mapped to the $Z$ Pauli operator (resp. $X$ Pauli operator) of the dual code that runs in the same direction.

One can identify the toric code and its dual very easily with physical qubit translation and relabeling. Doing so restores the original set of stabilizers, but the logical frame has changed: the original horizontal $X$ logical operator has been mapped to the horizontal $Z$ logical operator, which act on the other logical qubit. The logical operation implemented by the transversal Hadamard gate followed by the relabeling is thus a little more complex than a transversal logical Hadamard, as it also swaps the logical qubits \fig{fig:transHlog}.

Applying transversal Hadamard gates to a surface code with boundaries will change the type of the boundaries in addition of performing a logical Hadamard gate. This fact can be of importance in a context where both cylindrical patches and toric code are used: a transversal Hadamard on a \(Z\)-patch (resp. \(X\)-patch) will change it to a \(X\)-patch (resp. \(Z\)-patch), changing the allowed transversal two-qubit gates (see \secc{sec:sec:patch}).

Such an operation can be done in one time step, making it a constant time operation with respect to the distance of the code.

\subsection{\texorpdfstring{$S$}{S} gates and \texorpdfstring{$\ket{+_Y}$}{|+Y>} state preparation}
\label{sec:sec:stateprep}

It is possible to apply an $S$ gate simultaneously on both logical qubits of the toric code by folding it on itself, in a similar fashion to the one introduced in \cite{moussaTransversalCliffordGates2016}. One diagonal direction of the grid is chosen, and $S$ gates are applied along it. Using this diagonal as a symmetry axis, C$Z$ gates are applied from one side to the other. The square lattice is such that $X$-stabilizers are mapped to their product with a $Z$-stabilizer of the code while the $Z$ are unchanged. $X$ Pauli logical operators are multiplied by their $Z$ counterpart, while $Z$ Pauli logical operators are unchanged, hence implementing a logical transversal $S$ gate.\\

\begin{figure*}
  \centering
  \begin{subfigure}{0.45\linewidth}
    \centering
    \resizebox{0.75\linewidth}{!}{%
  \input{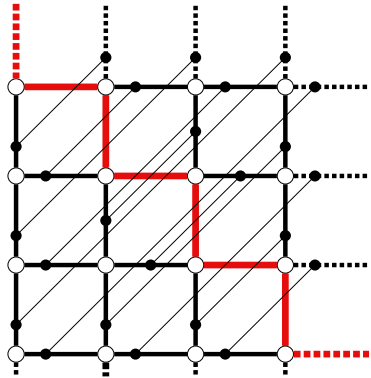}%

    }
  \end{subfigure}
  \caption{Simultaneous $S$ gates. Physical \(S\) gates are applied on the {\color{codered} colored} qubits, C$Z$ gates are applied symmetrically across.}
  \label{fig:prepy}
\end{figure*}

\noindent To prepare two copies of \ket{+_Y} states, one can proceed as follows:

\begin{enumerate}
\item Initialize every qubit of a toric lattice in the \ket{+} state
\item Measure every stabilizer of a toric code on said lattice (enough time to be confident in their measurement outcome)
\item Apply the folded $S$ gate procedure \fig{prepy}
\end{enumerate}

These are of particular interest as they can be used to implement single qubit logical $S$ gates (see \secc{sec:singles}). They could be prepared nonstop in ancillary block codes and used on the fly when needed.

This operation only requires one gate to be applied on every qubit, which can be performed in constant time with respect to the distance of the code.

\subsection{Single qubit Hadamard gate}

\begin{figure}
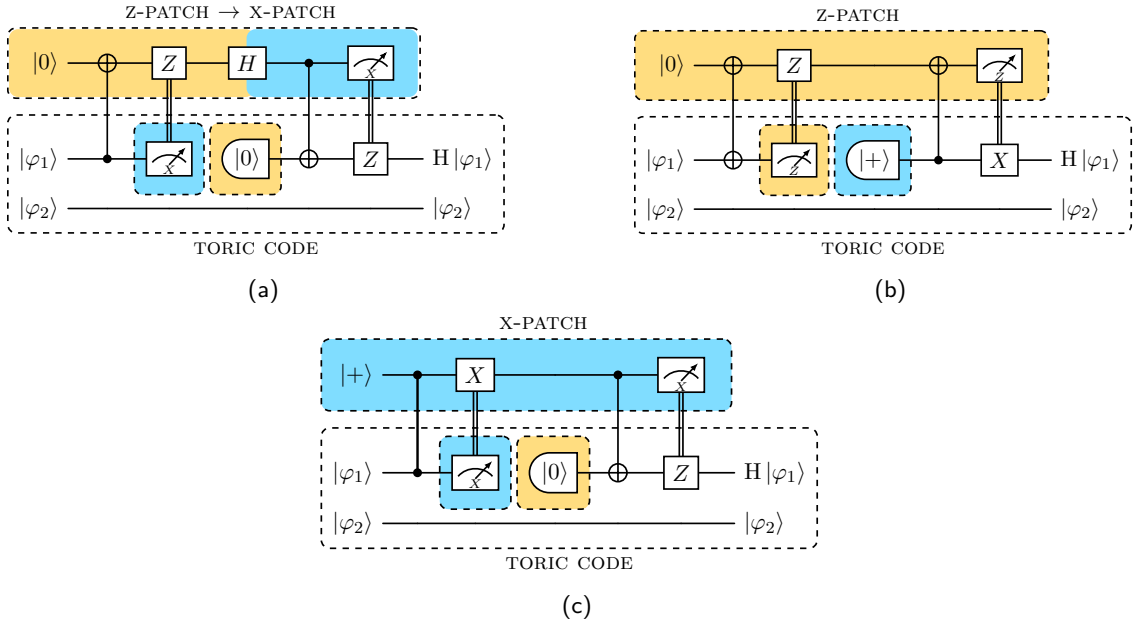

  \centering
  \begin{subfigure}{0.45\linewidth}
    \resizebox{\linewidth}{!}{%
  \input{./figs/singleH.tikz}%

    }
    \subcaption{}
    \label{fig:fig:singleH1}
  \end{subfigure}\hfill
  \begin{subfigure}{0.45\linewidth}
    \resizebox{\linewidth}{!}{%
  \input{./figs/singleH2.tikz}%

    }
    \subcaption{}
    \label{fig:fig:singleH2}
  \end{subfigure}

  \begin{subfigure}{0.45\linewidth}
    \resizebox{\linewidth}{!}{%
  \input{./figs/singleH3.tikz}%

    }
    \subcaption{}
    \label{fig:fig:singleH3}
  \end{subfigure}
  \caption{Protocols implementing a Hadamard gate on one logical qubit. (\subref{fig:fig:singleH1}) The qubit of interest is extracted, its basis is changed transversally in the patch and then reinserted back to torus. The boundaries of the patch are changed in the process. (\subref{fig:fig:singleH2}) The qubit of interest is extracted using a \XX~ gate which simultaneously change its basis. It can be then reinserted in the torus while keeping the same boundaries. (\subref{fig:fig:singleH3}) Same as (\subref{fig:fig:singleH2}), but uses a CZ and a \(X\)-patch.}
  \label{fig:hgate}
\end{figure}

The transversal Hadamard gate seen in \secc{sec:transH} is not sufficient in itself to achieve universal computation as it does not allow for logical single qubit Hadamard gate. We could isolate the logical qubit we want to apply the Hadamard gate on in a toric code by extracting the other qubit of the codeblock and then applying the transversal Hadamard gate. This approach is rather convoluted and would be hard to generalize to other block codes. Instead, we will be using the qubit displacement techniques developed in \secc{sec:logqbtp} to teleport the logical qubit we want to apply the Hadamard gate on from the block code to a cylindrical patch of surface code. The transversal Hadamard gate of the cylindrical code can then be used and the logical qubit would be inserted back to the torus \fig{fig:singleH1}.

Alternatively, the Hadamard gate can be performed for free during the qubit extraction or insertion by using two-qubit gates different than a CNOT gate (see Fig.\,\ref{fig:fig:singleH2} and \ref{fig:fig:singleH3}).

These procedures require the measurement of a single logical qubit of the torus, which requires an additional cylindrical patch of code (see \secc{sec:partmeas}). Since teleporting the logical qubit requires the target logical qubit to store a basis state, a reset has to be done on one logical of the torus, which again can be done with a partial measurement procedure followed by a classically controlled Pauli correction.

These procedures are just the chaining of fault-tolerant operations hence they are also fault-tolerant. These procedures use only constant time operations so they can be done in constant time, if the involved patches are prepared in advance as part of a larger computation.

\subsection{Single qubit \texorpdfstring{$\pi/2$}{pi/2} rotation gates}

\label{sec:sec:singles}

\begin{figure}
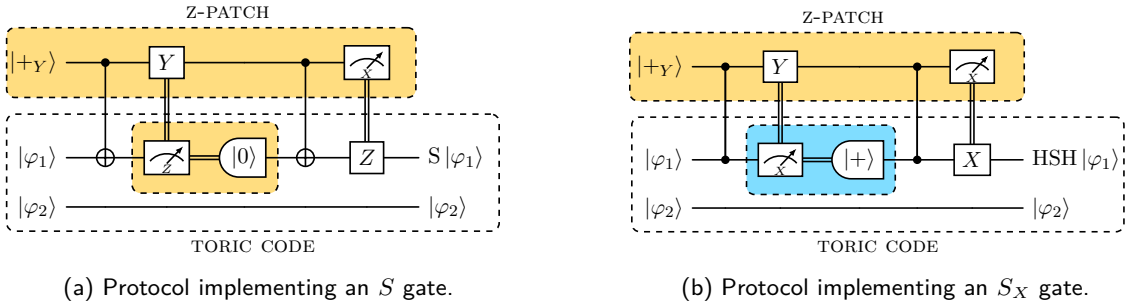

  \centering
  \begin{subfigure}{0.45\linewidth}
    \resizebox{\linewidth}{!}{%
  \input{./figs/singleS.tikz}%

    }
    \subcaption{Protocol implementing an $S$ gate.}
    \label{fig:fig:singleS}
  \end{subfigure}\hfill
  \begin{subfigure}{0.45\linewidth}
    \resizebox{\linewidth}{!}{%
  \input{./figs/singleS2.tikz}%

    }
    \subcaption{Protocol implementing an $S_X$ gate.}
    \label{fig:fig:singleS2}
  \end{subfigure}
  \caption{Protocols implementing $\pi$/2 rotation gates.}
  \label{fig:sgate}
\end{figure}

A $\frac{\pi}{2}$ rotation gate is required to generate the full Clifford group. Among this class of gates, the $S$ gate is particularly convenient Clifford gate as it can be used for implementing logical $T$ gates using some magic states \cite{bravyiUniversalQuantumComputation2005}. To perform an $S$ gate on a single logical qubit, we will once again use the qubit displacement techniques of \secc{sec:logqbtp} to teleport the logical qubit of a torus we want to act on into a cylindrical patch hosting a $\ket{+_Y}$ state. After a possibly needed $Y$ correction, the logical qubit is teleported back to the torus, which has been partially reset \fig{sgate}. The needed $\ket{+_Y}$ state can be obtained from a dedicated ancillary toric code on which the procedure from \secc{sec:stateprep} has been performed.

These procedures are just the chaining of five fault-tolerant operations hence they are also fault-tolerant. These procedures use only constant time operations so they can be done in constant time, if the involved patches are prepared in advance as part of a larger computation.

\subsection{$T$ gates}

\label{sec:sec:tgate}

\begin{figure*}
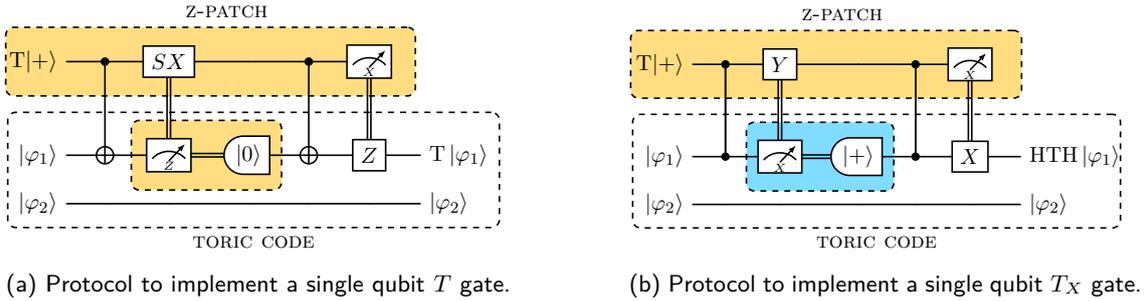

  \centering
  \begin{subfigure}{0.45\linewidth}
    \resizebox{\linewidth}{!}{%
  \input{./figs/Tdistill.tikz}%

    }
    \subcaption{Protocol to implement a single qubit $T$ gate.}
    \label{fig:fig:tgate}
  \end{subfigure}\hfill
  \begin{subfigure}{0.45\linewidth}
    \resizebox{\linewidth}{!}{%
  \input{./figs/Tdistill2.tikz}%

    }
    \subcaption{Protocol to implement a single qubit $T_X$ gate.}
    \label{fig:fig:tgate2}
  \end{subfigure}
  \caption{Fault-tolerant $\pi/4$ rotation gates implementation with a magic state.}
  \label{fig:TGateTP}
\end{figure*}

As is illustrated on Fig.~\ref{fig:TGateTP}, a single $T$ gate can be applied to a qubit by entangling a magic state with it and measuring it. Note that T gates can be transversally applied on a whole block code, provided the magic states are also stored in a block code.

This procedure is just the chaining of three fault-tolerant operations hence it is also fault-tolerant, provided the magic states are of sufficient quality. These procedures use only constant time operations so they can be done in constant time, if the involved patches and magic states are prepared in advance as part of a larger computation.

\section{Syndrome decoding and simulation results}
\label{sec:simulation}

Fault-tolerant properties of non-transversal gates can be challenging to demonstrate. In this section, we list a succession of numerical simulations exhibiting for each logical operation previously introduced the existence of a threshold for its logical error rate as well as the decoding procedure used.

Each experiment goes as follows: we start in a -- potentially faulty -- known logical state, perform the logical gate by applying faulty physical gates on and between physical qubits, repeatedly measuring the stabilizers at the end and sometimes during the logical gate, and finally measure all physical qubits to look for logical errors.
  
\subsection{Error model}
\label{sec:sec:errormodel}

Our study takes place in the phenomenological noise model: each single-qubit gate can introduce a random Pauli error $X$, $Y$ or $Z$ with equiprobability and each two-qubit gate can introduce any non-trivial Pauli error on two qubits with equiprobability. We allow for the measurement of any Pauli operator within a codeblock and without time constraints (stabilizers with intersecting support can be measured at the same time). These measurements are considered faulty (measure outcome flipped) and can introduce random single-qubit Pauli errors on each qubit of the measure support. While the support of some stabilizers might change during the computation, we try to keep measuring the same set of Pauli operators when possible and we restrain from measuring larger ones. All errors are mutually independent and occur with the same probability $\varepsilon$.

We will not consider encoding circuits in this paper. All simulated protocols start with a syndrome measurement step, whose role is to project the quantum state into the code space. When chaining logical gates during computation, it would coincide with the ending syndrome measurement step of the last applied gate.

\subsection{Decoding procedure and experiment description}
\label{sec:sec:decproc}

Logical error assessment is done using syndrome graph decoding. In the simplest case, all stabilizer measurements of an experiment are the vertices of a space-time graph, connected by one edge for each error (physical or from measurement) that can arise during computation and change the measurement outcome. Once errors have been sampled in any given run of a simulation, the syndrome is given to the decoding algorithm: stabilizers whose measurement outcome differs from the previous one are marked as belonging to the syndrome (they are often referred to as defects \cite{dennisTopologicalQuantumMemory2002} but \cite{gidneyStimFastStabilizer2021} calls them detection events). The decoding algorithm tries to infer the set of errors that occurred and decides whether some logical qubits have been flipped. The decoding procedure is deemed successful when inferred errors and sampled errors are in the same logical equivalence class. In this paper, we use \texttt{stim} as error-sampling framework, coupled with some specific decoders depending on the experiment.

\begin{figure*}
  \centering
  \resizebox{0.7\textwidth}{!}{%
  \input{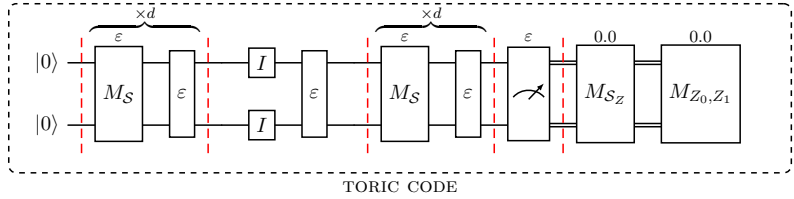}%

  }
  \caption{Memory experiment.}
  \label{fig:encod}
\end{figure*}

The horizontal strings represents logical qubits, here the two logical qubits of one toric code. The first step of Fig.\,\ref{fig:encod} means state preparation will ensure both logical qubits will be set to $\ket{0}$. The actual logical state will never impact the experiment, but has to be known for \texttt{stim} to define its logical observables. This state preparation step is considered perfect, but experiments will generally include $d$ rounds of stabilizer measurements to account for some errors in the state preparation.

The second and fourth steps are faulty syndrome extraction steps. The first box is the concurrent measurement of the Pauli operators from $\mathcal{S}$: the stabilizers of the code. The measurements are faulty and their classical outcome is flipped with a probability $\varepsilon$. The second box indicates some error locations: errors introduced by the physical operations immediately before will be sampled and taken into account starting from here by \texttt{stim} (see Sec.\,\ref{sec:sec:errormodel} for a precise description of the underlying quantum channels). Those errors occur with the same probability $\varepsilon$ for simplicity, but could be tuned to conform to specified hardware limitations. The boxes are wide as the involved processes can span the support of multiple representatives of logical Pauli operators of both logical qubits. 

The third step is a logical gate implementation step. The logical (transversal) identity gate is performed on the two qubits of the code. It is followed by an error location, which will introduce in this case depolarizing errors on all physical qubits with probability $\varepsilon$. We are describing a quantum memory that undergoes some noise before being read so we are not performing actual logical operations at this step, but merely introducing noise.

The fifth step represents the measurement of all physical qubits in the $Z$ basis. Measurement outcomes are flipped with probability $\varepsilon$, which accounts for both measurement and physical errors on these qubits.

The last step represents the classical reconstruction of the $Z$ stabilizers of the code (no measurement errors can be made at this stage: measurement errors from the single qubit measurements will be treated as qubit errors) and the reconstruction of the logical state measurements. As classical computation is supposed noiseless, the error rate of these two operations is set to 0.

\texttt{Stim} keeps track of deterministic observables and whether some sampled physical errors anticommute with them. A deterministic observable is a set of Pauli operator measurements whose product is predictable. Stabilizer measurements at one or more time steps can be seen as a deterministic observable. In general, for a given Clifford operator $U$, the measurement of any Pauli operator $P$ before the application of $U$ and the measurement of $UPU^{\dagger}$ afterwards form together a deterministic observable. All experiments will estimate the rate of occurrence of errors anticommuting with the defined observables. Usually, they will consist of the measurement of the initial logical state and a reconstructed logical measurement at the very end of the experiment. In this memory experiment, the two observables are the two $Z$ logical operators of the code measured at the beginning and reconstructed at the end, left untouched by conjugation by the identity operation performed.

\begin{figure}
  \centering
  \resizebox{\linewidth}{!}{%
    \includegraphics{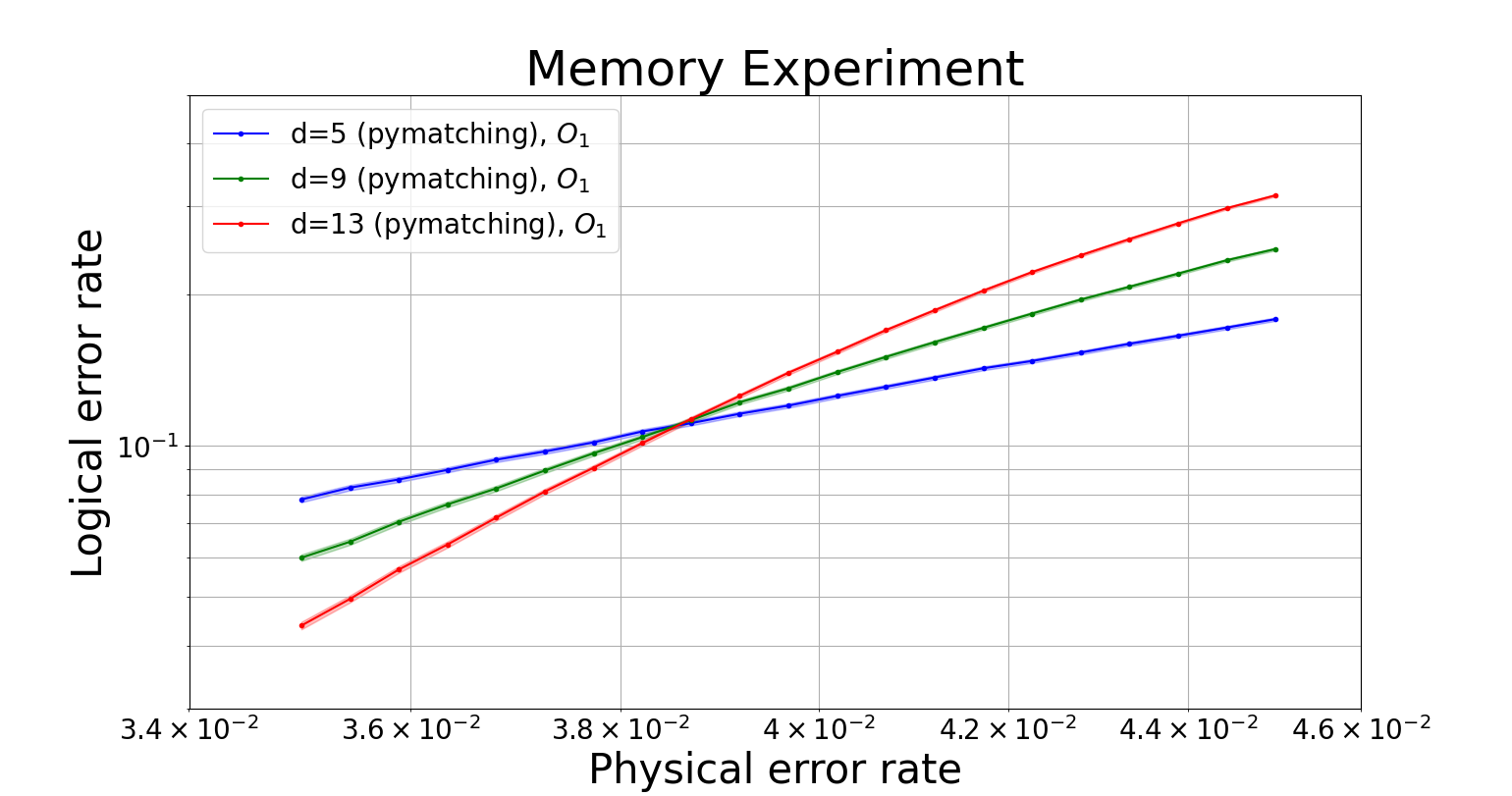}
  }
  \caption{Toric code memory logical error rate.}
  \label{fig:Memoryresults}
\end{figure}

\begin{figure}
  \centering
  \resizebox{\linewidth}{!}{%
    \includegraphics{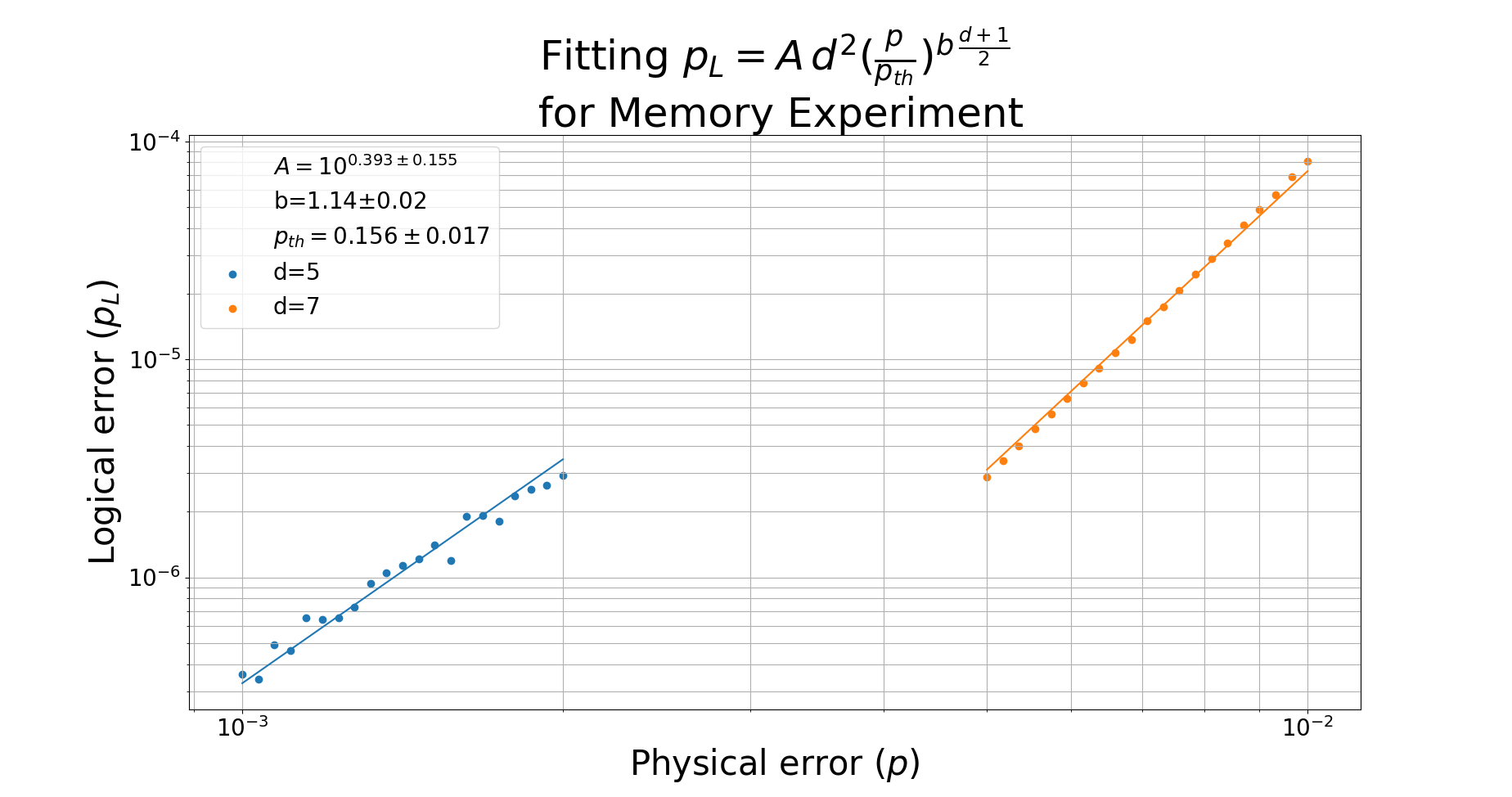}
  }
  \caption{Toric code memory logical error rate fitting for low physical error rate. The tracked observable \(O_1\) is \(Z_1\) at the beginning and end of the experiment.}
  \label{fig:fitmemory}
\end{figure}

This experiment only estimates the two $X/Y$ logical error rates of the logical operation \textit{i.e.} the rate at which an odd number of undetectable physical $X$ or $Y$ errors arises on the support of the observables. The experiment would have to be slightly modified to estimate the $Y/Z$ logical error rate. The Figure~\ref{fig:Memoryresults} shows a threshold at around 3.8\%. Experiments will also be run at low physical error rates, and fitted according to a model function \(f_{A, b, p_{th}}: p,d \mapsto Ad^{2} \left(\frac{p}{p_{th}}\right)^{b\frac{d+1}{2}}\) through a non-linear least square method. Parameter \(b\) is the effective distance reduction compared to the initial distance, \(p_{th}\) is the effective threshold at low physical error rate and \(A\) is scaling factor inherent to the specific studied protocol. For the memory experiment, a significantly higher effective threshold \(p_{th}\) of 15\% is found \fig{fitmemory}, together with an effective distance reduction \(b\) that shall be lower to 1. Simulating a memory experiment at low physical error rate is challenging, because it is hard to sample a logical error for \(d > 5\) and \(p < 10^{-2}\) which might explains this strange fit.

\begin{figure*}
  \centering
  \resizebox{0.95\linewidth}{!}{%
  \input{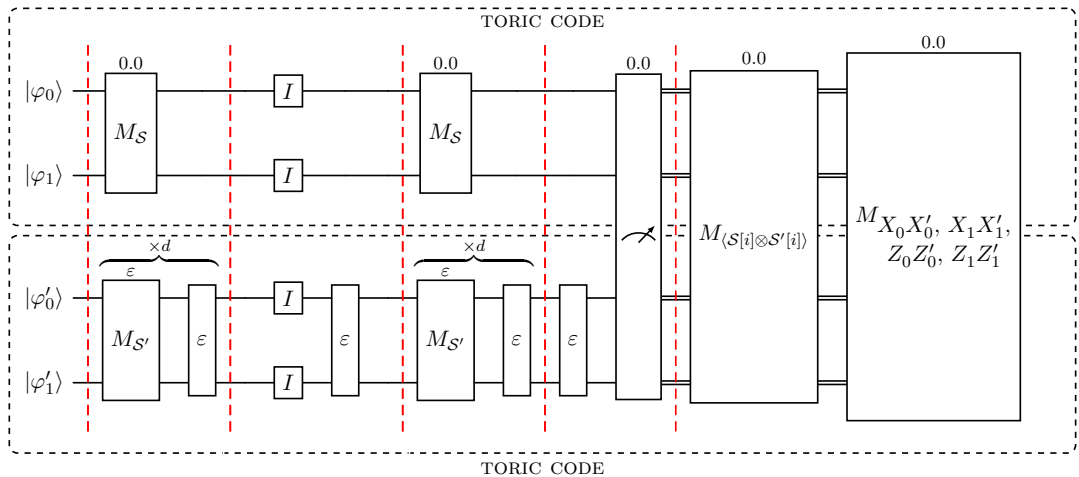}%

  }
  \caption{Complete memory experiment.}
  \label{fig:involvedmemory}
\end{figure*}

A more involved experiment can estimate all Pauli logical error rates in one go, at the expense of the use of a non-physical perfect copy of the experiment \fig{involvedmemory}.

We describe here how such an experiment would be built and represented for the dummy memory operation. We start with two Bell pairs as logical states shared by two toric codes: $\ket{\varphi_{0}\varphi_{0}^{\prime}} = \ket{\varphi_{1}\varphi_{1}^{\prime}} = \frac{\ket{00} + \ket{11}}{\sqrt{2}}$. One of these two codes is idealized: all measures and gate applications inside of it are perfect while the other is subjected to noise. The four logical operators $X_0X_{0}^{\prime}$, $X_1X_{1}^{\prime}$, $Z_0Z_{0}^{\prime}$ and $Z_1Z_{1}^{\prime}$ can be measured altogether, and can only be erroneous if a logical error occurred in the noisy code. They will be the tracked observables.

After the logical operation, the syndrome is perfectly measured in the ideal code and imperfectly measured in the noisy one. One key difference with the previous setup is the ending physical measurement of qubits. To be able to infer the value of all four logical operators at once, physical qubits of both codes have to be measured jointly (\textit{i.e.} by pair) in both basis. These measures have to be perfect to affirm that no logical errors at the end come from the ideal code. To compensate for the use of this non-physical operation, an additional error location is added just before in the noisy code. At last, stabilizers measurement outcomes and logical operators values are jointly inferred from the physical measurements and fed to the decoder. This gives us estimates for the error rate of the four observables, the $X$ and $Z$ observables of either logical qubit.

\subsection{Transversal CNOT gate}
\label{sec:sec:TCNOT}

The transversal CNOT gate is already known to be fault-tolerant by definition. However, estimating its logical error rate raises some interesting difficulties that also come up in the estimation of other error rates.

\begin{figure}
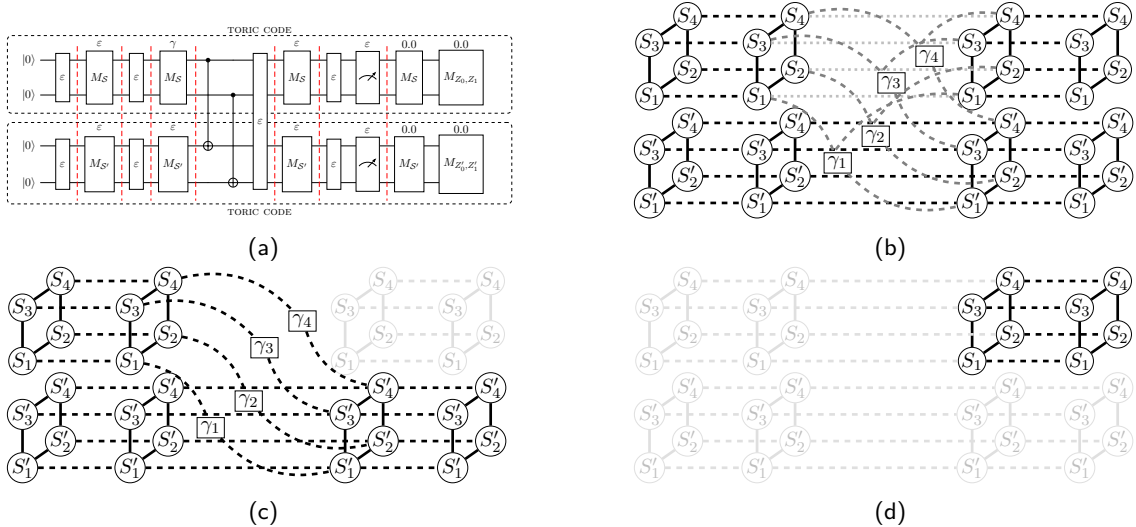

  \centering
  \begin{subfigure}{0.45\linewidth}
    \resizebox{\linewidth}{!}{%
  \input{./figs/dummy2toriccircuit.tikz}%

    }
    \subcaption{}
    \label{fig:hyperg:exp}
  \end{subfigure}
  \hfill
  \begin{subfigure}{0.45\linewidth}
    \resizebox{\linewidth}{!}{%
  \input{./figs/dummy2toricdetectors.tikz}%

    }
    \subcaption{}
    \label{fig:hyperg:graph}
  \end{subfigure}

  \begin{subfigure}{0.45\linewidth}
    \resizebox{\linewidth}{!}{%
  \input{./figs/dummy2torictwo-step.tikz}%

    }
    \subcaption{}
    \label{fig:hyperg:two-step}
  \end{subfigure}\hfill
  \begin{subfigure}{0.45\linewidth}
    \resizebox{\linewidth}{!}{%
  \input{./figs/dummy2torictwo-step2.tikz}%

    }
    \subcaption{}
    \label{fig:hyperg:two-step2}
  \end{subfigure}
  \caption{Transversal CNOT experiment and associated decoding graph.}
  \label{fig:hyperg}
\end{figure}

Consider the experiment depicted on Figure~\ref{fig:hyperg:exp} where a transversal CNOT gate is performed between two distance 2 toric codes. We only estimate the $X$ logical error rate to simplify the discussion. The states are prepared and their stabilizers measured before the transversal CNOT gates are applied (we represent them as two CNOT between the logical qubits, as it is their logical effect on the logical qubits). We then measure the \emph{same} stabilizer support before the physical measurements and stabilizer reconstructions. The corresponding decoding graph is shown on Figure~\ref{fig:hyperg:graph}. The vertices are the $Z$ stabilizers of both codes, measured at different time steps (step 2, 4, 6 and 8 of the experiment). Plain edges are qubit defects. They only appear at four time steps, corresponding to the error locations of step 1, 3, 5 and 7 of the experiment. They flip only two spatially adjacent detection events, as their effect on the measurement outcomes of those stabilizers is permanent. Dashed edges are stabilizer measurement errors : the outcome has incorrectly be read as -1 instead of 1 for instance. Most of them involve only one stabilizer but two detection events as their effect on the outcome is not permanent : the following measurement will differ from the previous one in the case where no other error occurs. Measurement errors happen only at three different time steps, as the last measurement outcomes are reconstructed from the measured physical qubits. The four measurement errors linked to the error location with rate $\gamma$ are hyperedges instead of regular edges. Since we do not change the support of the stabilizers we measure after the CNOT, we in fact change the set of stabilizers we measure. To be consistent, the expected value of these stabilizers is a combination of the values of some stabilizers measured previously. A measurement error before the CNOT gates have repercussions on more than one detection event immediately after them if its outcome is used in several stabilizer recombinations. In Figure~\ref{fig:hyperg:graph}, the four $Z$-stabilizers of the second toric code are mapped by the CNOT gates to their product with their counterpart in the first code. Hence, measurement errors of the four $Z$-stabilizers of first code at location $\gamma$ not only flip the next detection event linked to their measurement but also the detection event linked to the measurement of one $Z$-stabilizer of the second code. The dotted edges are the measurement error edges that would replace the hyperedges had we allowed ourselves to change (and consequently increase the weight of) the measured stabilizers.

The presence of hyperedges in the decoding graph is a consequence of the objective to keep measuring low-weight stabilizers. An experiment with hyperedges in its decoding graph cannot be directly decoded using the Minimal Weight Perfect Matching decoder, as finding a perfect matching in a hypergraph is NP-hard (the associated decision problem has been proven NP-complete by \cite{Karp1972}). In some cases, hyperedges can be split into regular edges whose dependency is forgotten. Some heuristics achieving this exist in \texttt{stim}, but they mainly rely on decomposed edges to be already present in the decoding graph (\textit{e.g.} by decomposing $Y$ errors into $Z$ and $X$ errors), which is rarely the case when we do stabilizer recombinations. Instead, we use two different decoders to decode our experiments when some hyperedges appear in the decoding graph.

The first decoder is the Hierarchical Minimal Weight Perfect Matching decoder (or two-step decoder). Given a syndrome, we first focus on one of the two codes and starting from the last time step, we select all ancestor of the detection events of this code \fig{hyperg:two-step}. We solve the MWPM problem in the induced decoding graph considering only detection events from the syndrome still present in the induced graph. We then remove from the syndrome defects that have been dealt with by the first matching. If some errors at the $\gamma$ location have been inferred in the first matching, we add the third vertex from the corresponding hyperedges to the syndrome. We then compute a second matching considering only detection events in the first code \fig{hyperg:two-step2} considering the updated syndrome.

This decoding method always find a valid correction for the syndrome, but has several drawbacks:
\begin{itemize}
\item It is significantly slower than the current MWPM decoder available through \texttt{stim}, as we need to solve two matching problems but also because we have to retrieve the actual errors inferred by the first step to correctly update the syndrome for the second step instead of just keep track of the logical Pauli frame.
\item It can only be used as is for small depth experiments. Indeed, to start the second decoding, we need the result of the first decoding, which only become available when the physical qubits of the second code are individually measured. This is not a problem as we are only considering isolated logical operations, but chaining entangling gates between two or more codeblocks would require further refinements of this method.
\end{itemize}

This decoding strategy is very similar to the MWPM approach used in \cite{cainCorrelatedDecodingLogical2024} to decode a similar experiment. It also performs two consecutive matchings to tackle the hyperedge appearance. One difference is that our decoder splits the checks spatially as well as according to time, leading to 75\% of the syndrome being available for the first matching.

The second decoder we considered is the so-called Union-Find decoder which grows clusters of detection events containing uncorrectable defects until they become correctable and returns a set of errors. Using this algorithm directly on the decoding hypergraph seems to be enough in most tested cases (with enough code distance) to exhibit a threshold. This approach has also its own drawbacks:

\begin{itemize}
\item The decoding threshold of the UF decoder is below the one of the MWPM decoder \cite{delfosseAlmostlinearTimeDecoding2021}
\item Having hyperedges in the decoding graph negates most computational advantages of this method: even-weight clusters are not necessarily correctable anymore, some odd-weight clusters are. A check step has to be performed after every growing steps, which has a dramatic impact on the complexity 
\end{itemize}

\begin{figure}
  \centering
  \begin{subfigure}{\linewidth}
    \resizebox{\linewidth}{!}{%
      \includegraphics{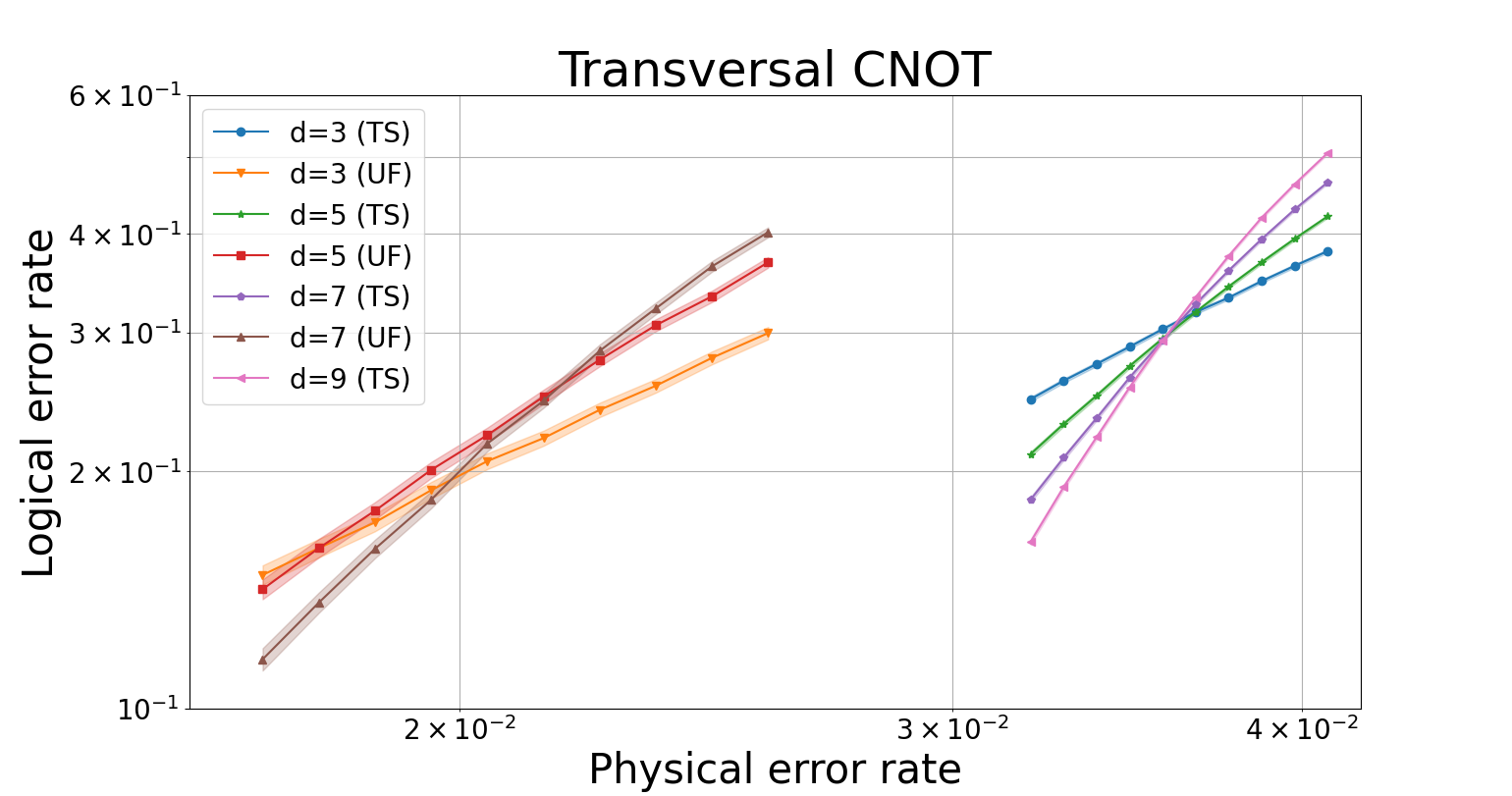}
    }
  \end{subfigure}
  \caption{Comparison between estimated logical error rate with the two heuristics: the \texttt{two-step} decoder (TS) and the UnionFind decoder (UF). The tracked observables are \(X_0\), \(X_1\), \(X_0^{\prime}\) and \(X_1^{\prime}\) becoming respectively \(X_0X_0^{\prime}\), \(X_1X_1^{\prime}\), \(X_0^{\prime}\) and \(X_1^{\prime}\) in the end. A logical error is counted when any observable is flipped.}
  \label{fig:transCNOTresults}
\end{figure}

The final step's reconstructed logical measurements confronts respectively initial $Z_0$, $Z_1$, $Z^{\prime}_0$ and $Z^{\prime}_1$ with ending $Z_0$, $Z_1$, $Z_0Z^{\prime}_0$ and $Z_1Z^{\prime}_1$ to form the four observables. The Figure \ref{fig:transCNOTresults} shows the difference between the two custom decoders for the estimation of the combined logical error rate. The \texttt{two-step} procedure uncovers a neat 3.6~\% threshold, while the \texttt{UFDecoder} sketches a threshold at around 2~\%. For experiments below, in the case the regular MWPM decoder offered by \texttt{stim} and \texttt{pymatching} couldn't be used, we will state the decoder we used to obtain the logical error rate plots.

\subsection{Non-transversal CNOT gates}
\label{sec:sec:CNOTexp}

\subsubsection{Dehn twist simulation}
\label{sec:sec:sec:DTexp}

\begin{figure}
  \centering
  \resizebox{\linewidth}{!}{%
    \includegraphics{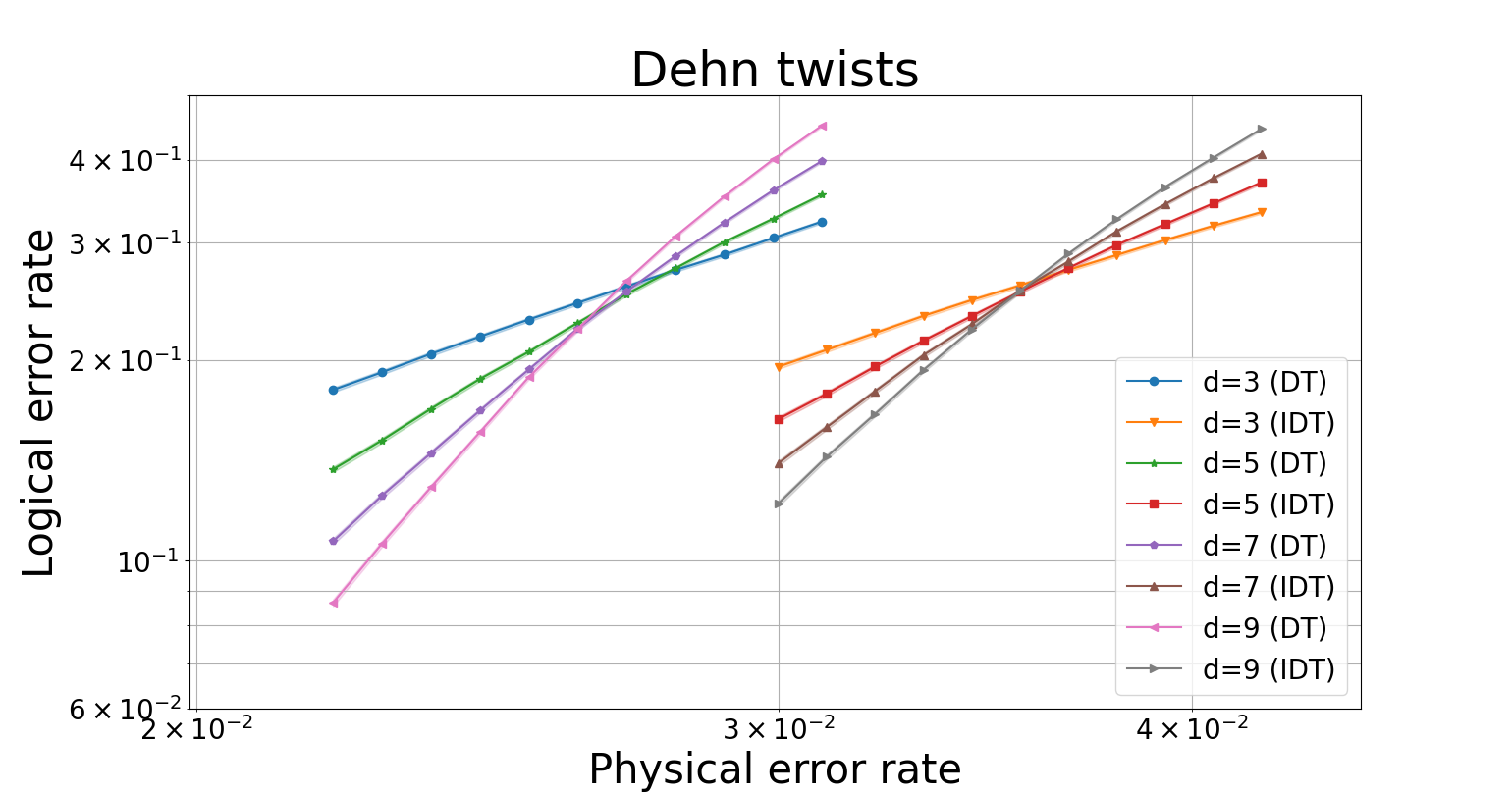}
  }
  \caption{Logical error rate estimations of the Dehn twist operation (DT) and instantaneous Dehn twist operation (IDT) using \texttt{pymatching}. The tracked observables are \(Z_0\) and \(Z_1\) becoming respectively \(Z_0\) and \(Z_0Z_1\) in the end. A logical error is counted when either or both observables are flipped.}
  \label{fig:DTresults}
\end{figure}

The Dehn twist simulation protocol follows details explained in \secc{sec:sec:DT}. Between logical state initialization and physical qubit measurements, $d$ Dehn twists steps are performed: noisy CNOT gates along representatives of two logical operators are performed, immediately followed by a round of stabilizer measurements from the set of stabilizers. As stabilizer operators only change support but are not combined, the usual \texttt{pymatching} decoding can be used. The final step's reconstructed logical measurements confronts respectively initial $Z_0$ and $Z_1$ with ending $Z_0$ and $Z_0Z_1$ to form the two observables. We obtain a threshold of 2.7~\% \fig{DTresults}.

\begin{figure*}
  \centering
  \resizebox{\textwidth}{!}{%
  \input{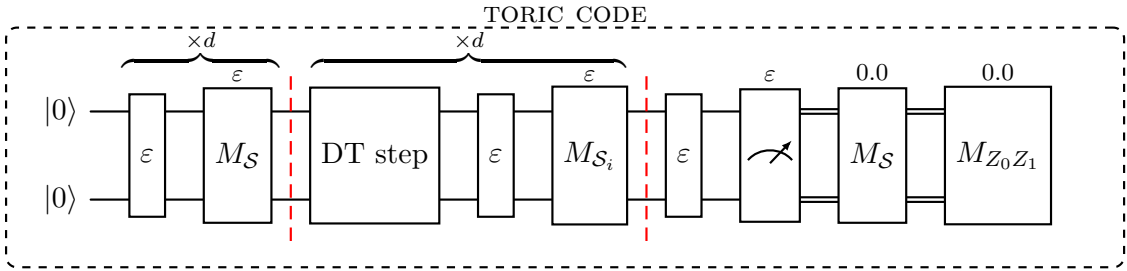}%

  }
  \caption{Dehn twist simulation protocol. $\mathcal{S}_{i}$ is the set of measured stabilizers at iteration $i$. $\mathcal{S}_{0}$ = $\mathcal{S}_{d}$ = $\mathcal{S}.$}
  \label{fig:DTexp}
\end{figure*}

\subsubsection{Instantaneous Dehn twist simulation protocol}
\label{sec:sec:sec:IDTexp}

\begin{figure}
  \centering
  \resizebox{\linewidth}{!}{%
    \includegraphics{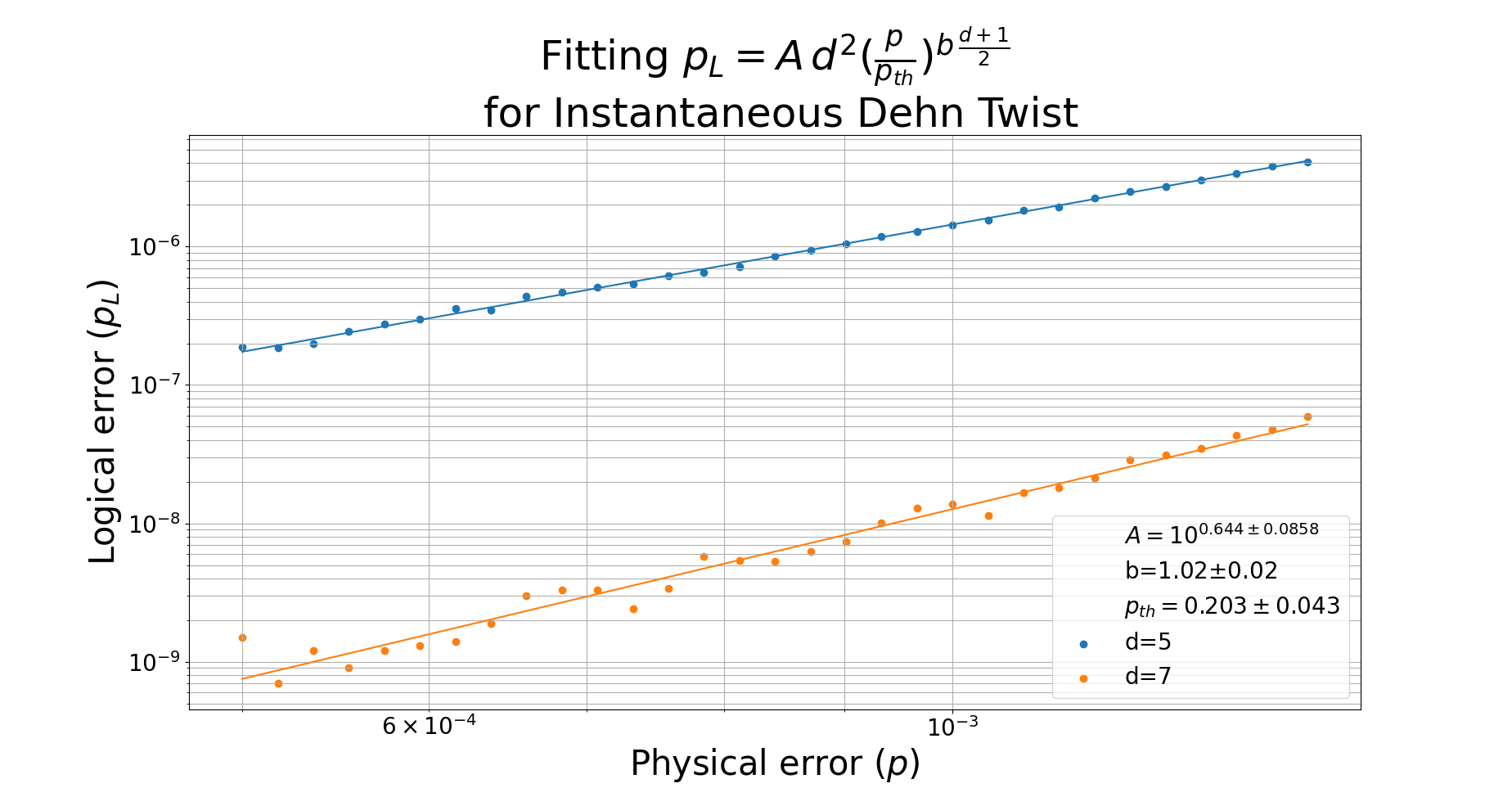}
  }
  \caption{Instantaneous Dehn twist marginal logical error rate fitting for low physical error rate.}
  \label{fig:fitIDT}
\end{figure}

The instantaneous Dehn twist simulation protocol follows details explained in \secc{sec:sec:IDT}. Between logical state initialization and physical qubit measurements, one step with noisy CNOT gates around the whole code is performed, durably changing the set of measured stabilizers. As stabilizer operators only change support but are not combined, the usual \texttt{pymatching} decoding can be used. The final step's reconstructed logical measurements confronts respectively initial $Z_0$ and $Z_1$ with ending $Z_0$ and $Z_0Z_1$ to form the two observables. We obtain a threshold of 3.5~\% \fig{DTresults}. While the error rate at threshold is a bit higher for the instantaneous Dehn twist than for the Dehn twist operation, the logical error rate of the instantaneous Dehn twist is smaller at a given physical error rate (below its threshold).
The fit along model function \(f_{A, b, p_{th}}\) for the instantaneous Dehn twist protocol at low physical error rate features an effective distance reduction \(b\) close to 1 \fig{fitIDT}, indicating that code distance wasn't impacted during the procedure. The effective threshold (20\%) is still highly optimistic compared to the results shown on Figure~\ref{fig:DTresults}.

\begin{figure*}
  \centering
  \resizebox{\textwidth}{!}{%
  \input{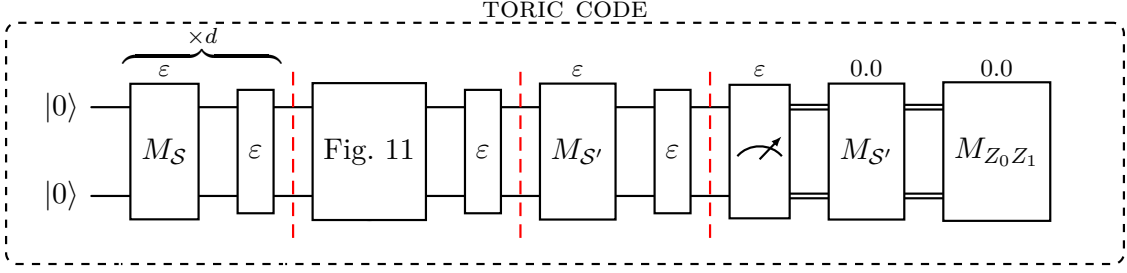}%

  }
  \caption{Instantaneous Dehn twist simulation protocol.}
  \label{fig:IDTexp}
\end{figure*}

\subsection{CZ gate}
\label{sec:sec:sec:CZexp}

\begin{figure*}
  \centering
  \resizebox{!}{5cm}{%
  \input{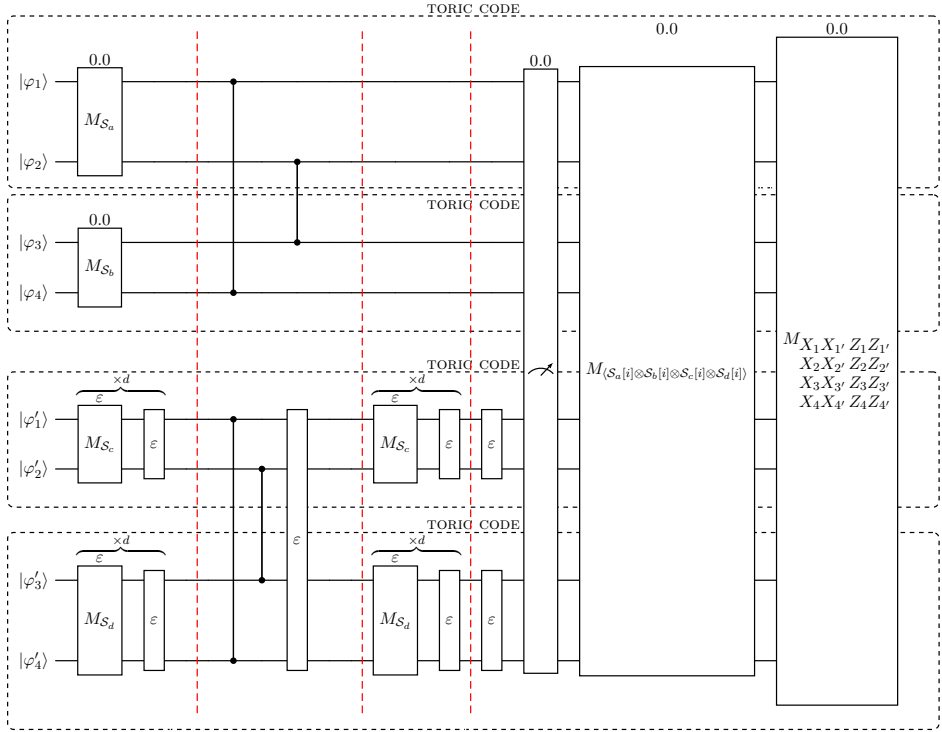}%

  }
  \caption{Transversal CZ gates simulation protocol.}
  \label{fig:CZexp}
\end{figure*}

\begin{figure}
  \centering
  \resizebox{\linewidth}{!}{%
    \includegraphics{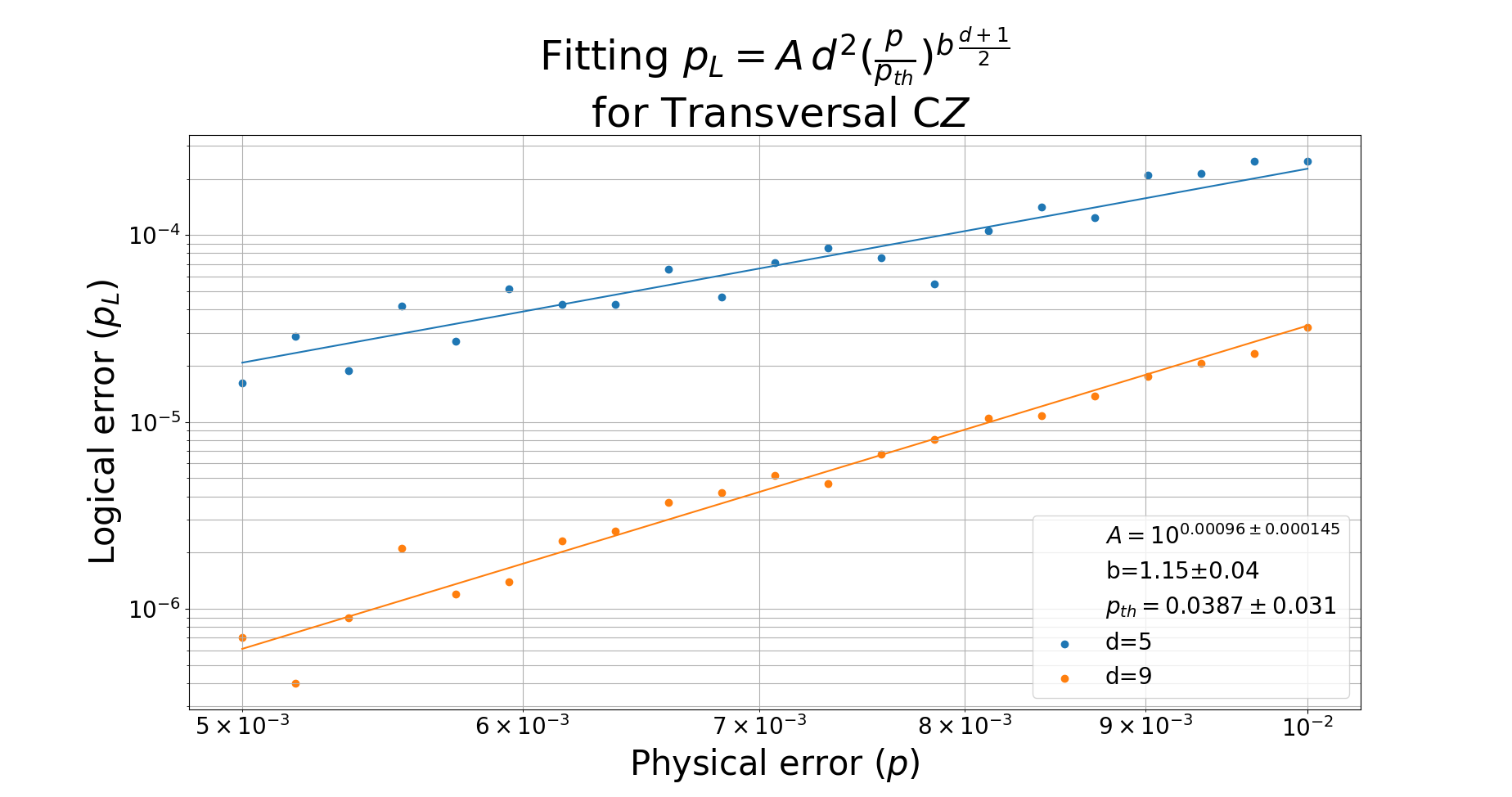}
  }
  \caption{Transversal CZ marginal logical error rate fitting for low physical error rate.}
  \label{fig:fitCZ}
\end{figure}

The transversal CZ logical error rates are estimated through the protocol described on Figure~\ref{fig:CZexp}. We start with two physical noisy toric codes and two ideal noiseless toric codes. Qubit states are entangled by pairs such that one qubit of each pair lies in a noisy code and the other in a noiseless code $\left(\ket{\varphi_{1}\varphi_{1}^{\prime}} = \ket{\varphi_{2}\varphi_{2}^{\prime}} = \ket{\varphi_{3}\varphi_{3}^{\prime}} = \ket{\varphi_{4}\varphi_{4}^{\prime}} = \frac{\ket{00} + \ket{11}}{\sqrt{2}}\right)$. The final step's reconstructed logical measurements confronts respectively initial $X_1X^{\prime}_1$, $X_2X^{\prime}_2$, $Z_1Z^{\prime}_1$, $Z_2Z^{\prime}_2$, $X_3X^{\prime}_3$, $X_4X^{\prime}_4$, $Z_3Z^{\prime}_3$ and $Z_4Z^{\prime}_4$ with ending $X_1X^{\prime}_1Z_4Z^{\prime}_4$, $X_2X^{\prime}_2Z_3Z^{\prime}_3$, $Z_1Z^{\prime}_1$, $Z_2Z^{\prime}_2$, $X_3X^{\prime}_3Z_2Z^{\prime}_2$, $X_4X^{\prime}_4Z_1Z^{\prime}_1$, $Z_3Z^{\prime}_3$ and $Z_4Z^{\prime}_4$ to form the eight observables. 
Interestingly such a protocol allows to access the correlation between logical errors, see Appendix~\ref{app:corr}.
We obtain a threshold of 3.6~\% \fig{CZresults}.
The fit along model function \(f_{A, b, p_{th}}\) for this transversal  C\(Z\) protocol at low physical error rate features an effective distance reduction \(b\) greater than 1 \fig{fitCZ}, just like for the memory experiment. This protocol should be investigated at even lower error rate, which is very CPU-time consuming given the size of the circuit and the overall low probability of logical error.

\begin{figure}
  \centering
  \resizebox{\linewidth}{!}{%
    \includegraphics{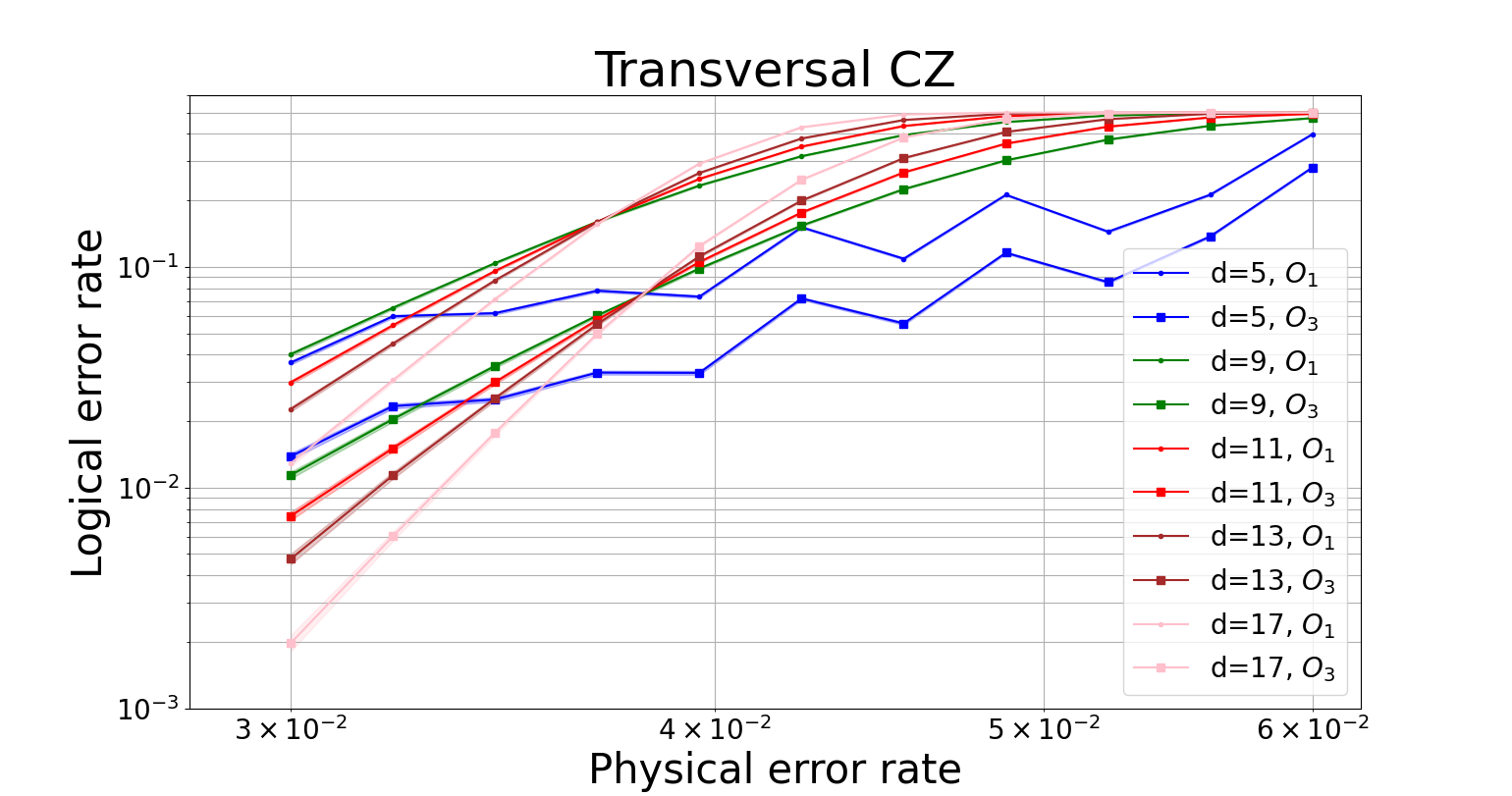}
  }
  \caption{Logical error rate estimation of the transversal CZ gate operation using \texttt{two-steps pymatching}. $Z/Y$ logical error rate ($O_1$) is compared to $X/Y$ logical error rate ($O_3$).}
  \label{fig:CZresults}
\end{figure}

\begin{figure}
  \centering
  \resizebox{\linewidth}{!}{%
    \includegraphics{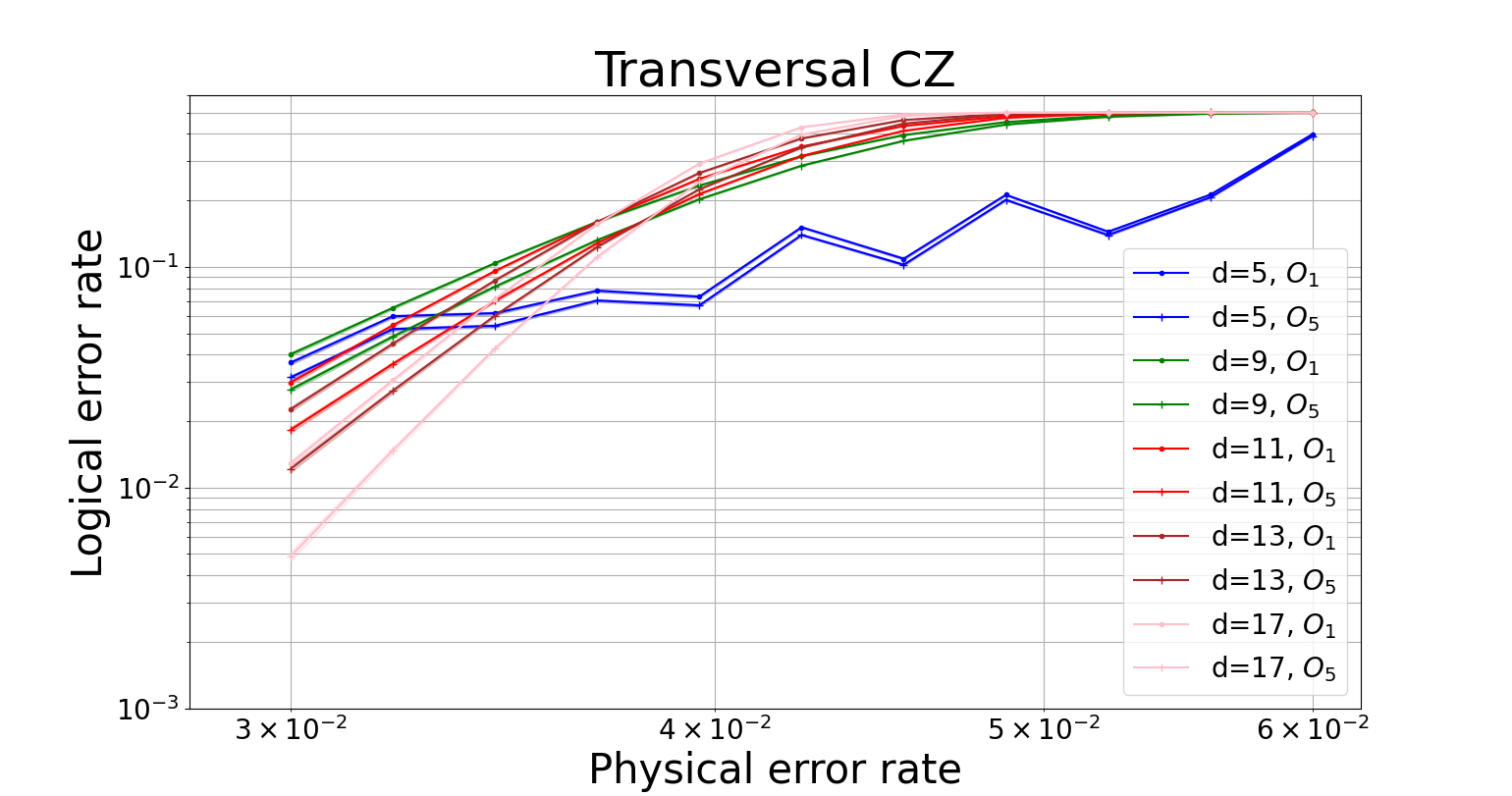}
  }
  \caption{Logical error rate estimation of the transversal CZ gate operation using \texttt{two-steps pymatching}. $Z/Y$ logical error rate of the one logical qubit of a block ($O_1$) is slightly worse than the one from the other block ($O_5$). This shows that the \texttt{two-steps pymatching} decoder is better at decoding of block than the other.}
  \label{fig:CZresults}
\end{figure}

\subsection{Transversal Hadamard gate}
\begin{figure*}
  \centering
  \resizebox{\textwidth}{!}{%
  \input{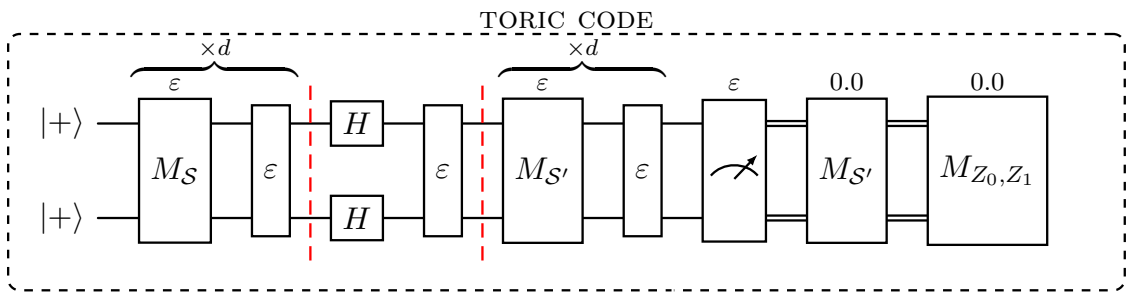}%

  }
  \caption{Transversal Hadamard simulation protocol.}
  \label{fig:transHexp}
\end{figure*}

\begin{figure}
  \centering
  \resizebox{\linewidth}{!}{%
    \includegraphics{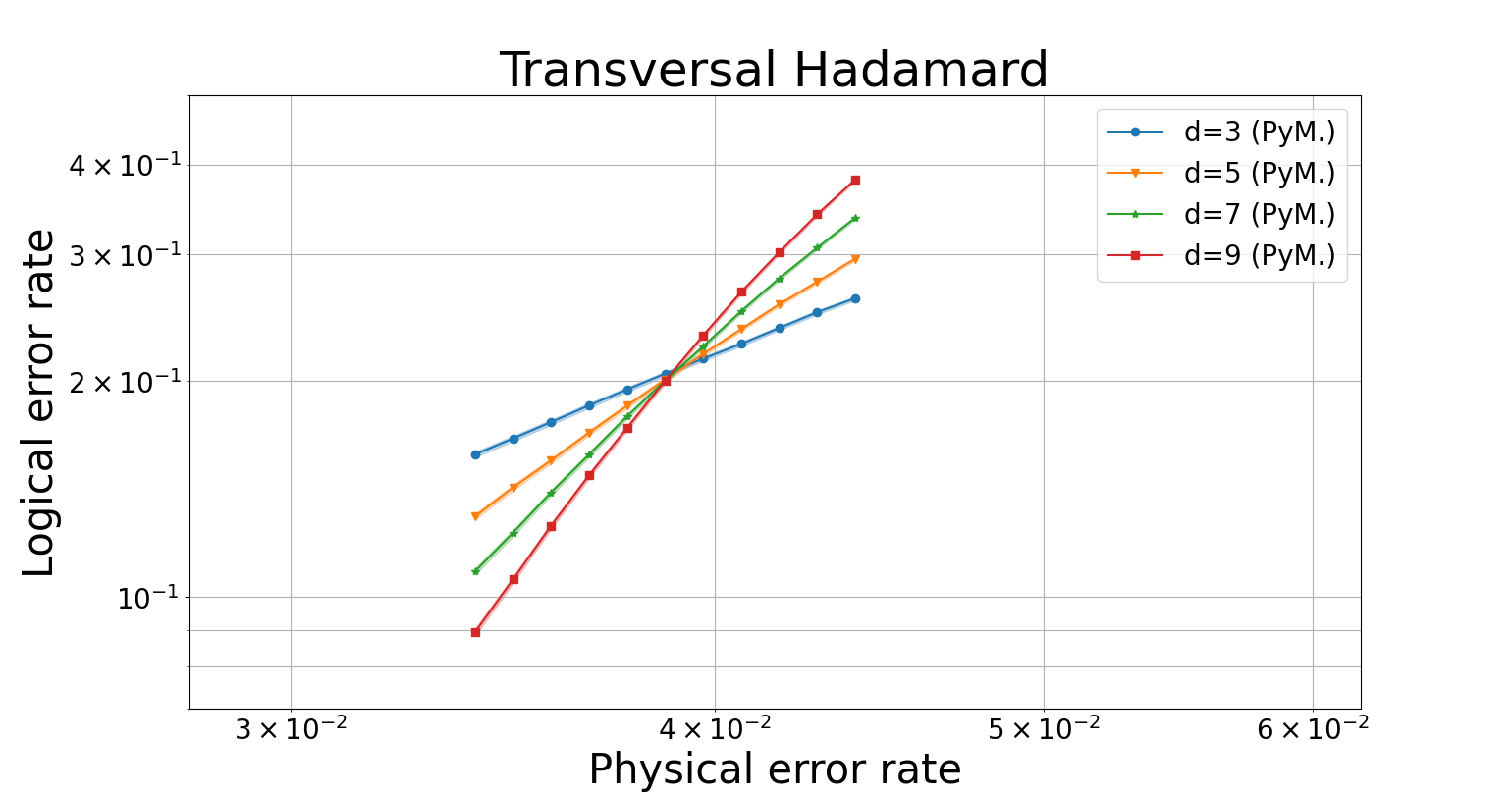}
  }
  \caption{Logical error rate estimation of the transversal \(H\) gate operation using \texttt{pymatching}. The tracked observables are \(Z_0\) and \(Z_1\) becoming respectively \(X_0\) and \(X_1\) in the end. A logical error is counted when either or both observables are flipped.}
  \label{fig:transHresults}
\end{figure}

The transversal \(H\) logical error rates are estimated through the protocol described on Figure~\ref{fig:transHexp}. Between logical state initialization and physical qubit measurements, one step with noisy Hadamard gates around is performed transversally, durably changing the set of measured stabilizers. This change is involutive which means it does not impact much the required connectivity. The final step's reconstructed logical measurements confronts respectively initial $X_0$ and $X_1$ with ending $Z_0$ and $Z_1$ to form the two observables. We obtain a threshold of 3.8~\% \fig{transHresults}.

\subsection{Partial measurement}
\begin{figure*}
  \centering
  \resizebox{\textwidth}{!}{%
  \input{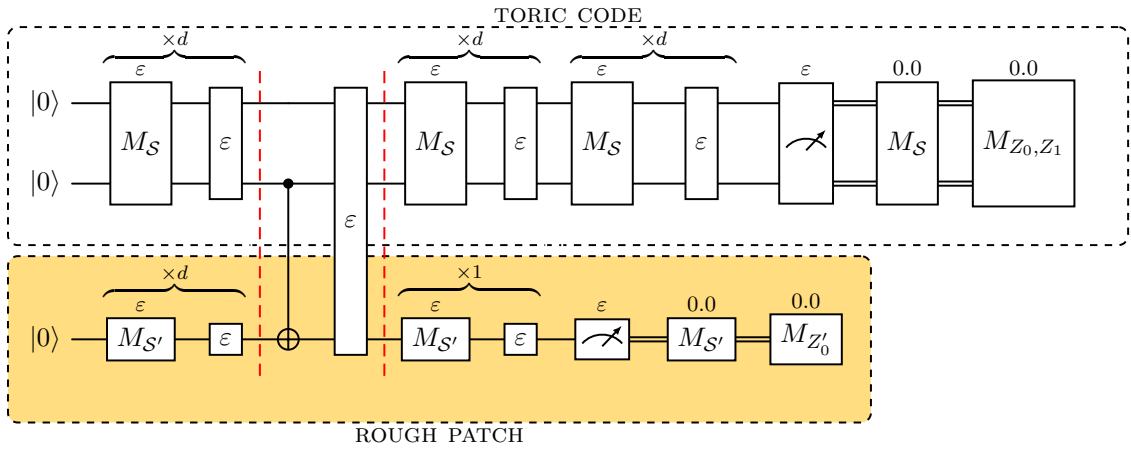}%

  }
  \caption{Partial measurement protocol.}
  \label{fig:partmeas}
\end{figure*}

\begin{figure}
  \centering
    \resizebox{\linewidth}{!}{%
      \includegraphics{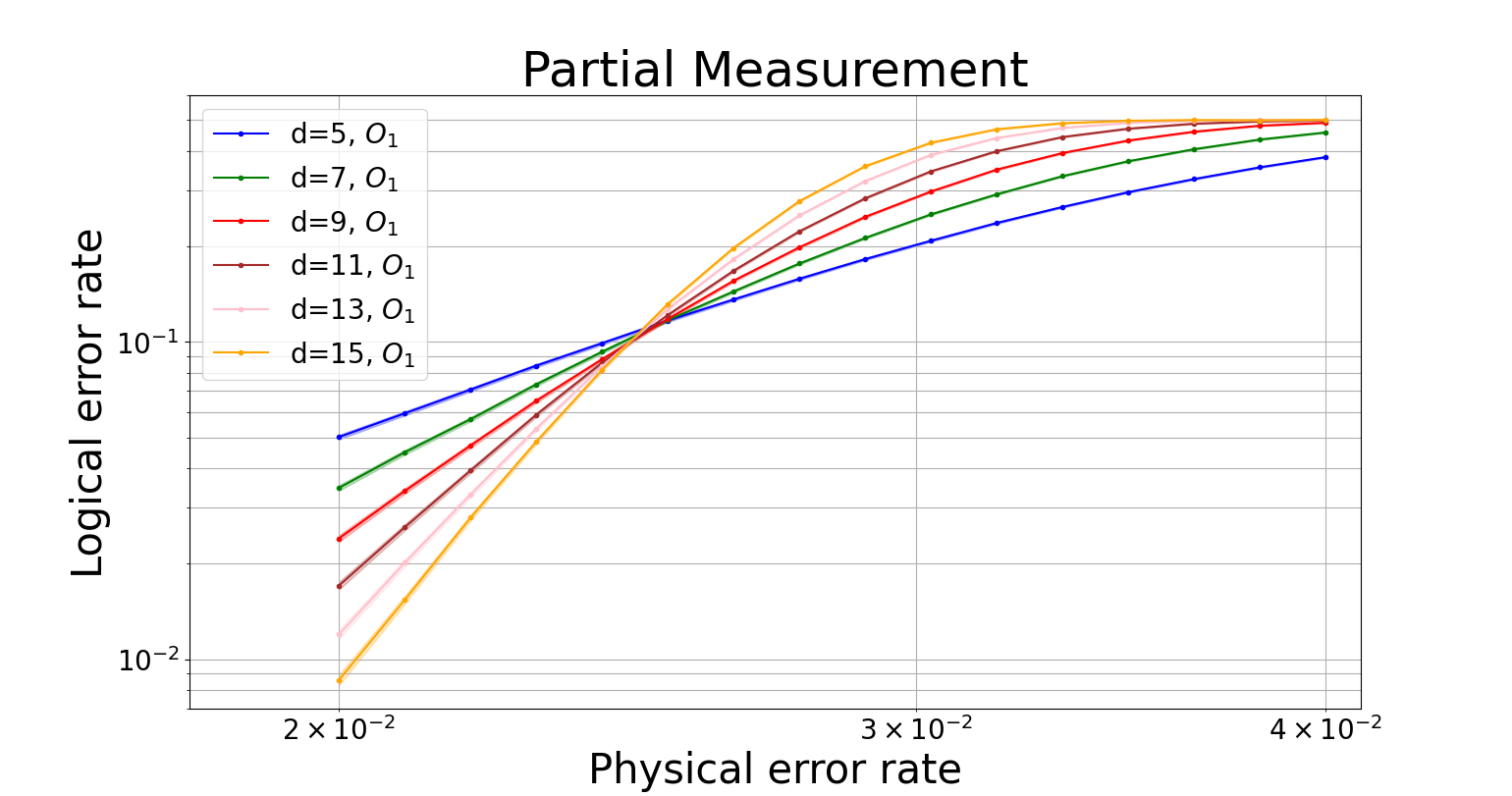}%
    }
    \caption{Logical error rate estimation of the partial measurement operation using \texttt{two-step pymatching}. The tracked observables are \(Z_0\) and \(Z_1\) that should stay the same throughout the measurement. The two error rate are very similar so only the first is depicted for clarity.}
    \label{fig:PMlog}
  \end{figure}
  \begin{figure}
    \resizebox{\linewidth}{!}{%
      \includegraphics{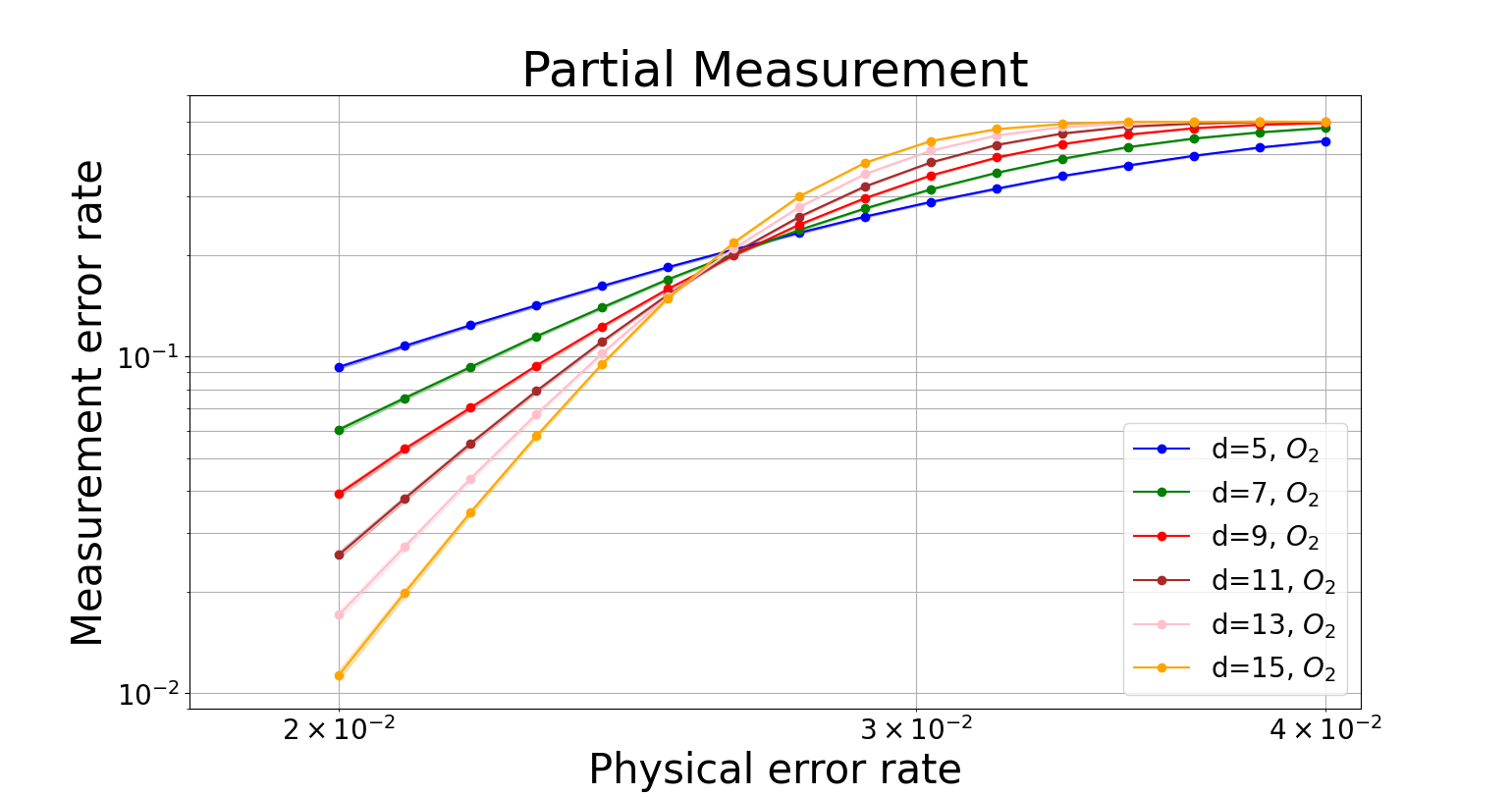}%
    }
  \caption{Measurement error rate estimation of the partial measurement operation using \texttt{two-step pymatching}. The tracked observable is \(Z_1Z_0^{\prime}\) becoming \(Z_0^{\prime}\) after the measurement.}
  \label{fig:PMmeas}

\end{figure}

\begin{figure}
  \centering
  \resizebox{\linewidth}{!}{%
    \includegraphics{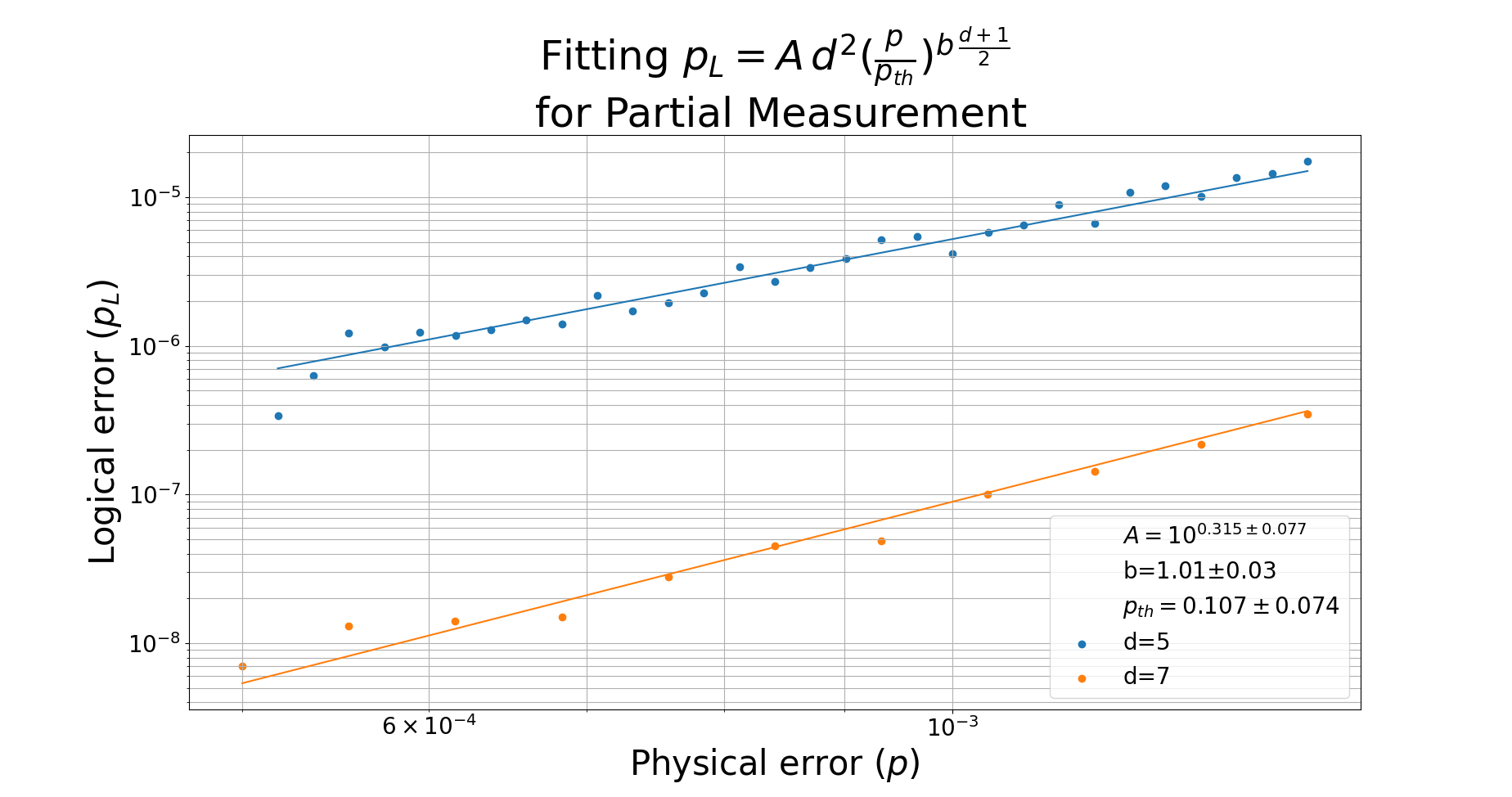}
  }
  \caption{Partial Measurement measurement error rate fitting for low physical error rate.}
  \label{fig:fitPM}
\end{figure}

Both the logical error rate and logical measurement error rate are estimated through the protocol described on Figure~\ref{fig:partmeas}.
We start with a noisy toric code and one rough patch initialized in the $\ket{0}$ state. After logical state initialization, some transversal CNOT gates between the toric code and the patch are performed, as per \secc{sec:partmeas} alongside representatives of logical qubit 1.
The patch is then measured, while the code undergoes other operations, here an identity operations, before its physical qubits are also measured.
The final step's reconstructed logical measurements confronts respectively initial $Z_0$, $Z_1$ with ending $Z_0$, $Z_1$ to form two observables tracking the logical error rate of the procedure, while the initial $Z^{\prime}_0$ is confronted with ending $Z^{\prime}_0Z_1$ to form one observable tracking the measurement error rate of the procedure.
We obtain a threshold of 2.4~\% for the logical qubit error rate \fig{PMlog} and a threshold of 2.5~\% for the measurement error rate \fig{PMmeas}.
The fit along model function \(f_{A, b, p_{th}}\) for the partial measurement protocol at low physical error rate features an effective distance reduction \(b\) close to 1 \fig{fitPM}, indicating that code distance wasn't impacted during the procedure. The effective threshold (10\%) is still highly optimistic compared to the results shown on Figure~\ref{fig:IDTexp}.

\subsection{Folded \(S\) gate}
\begin{figure*}
  \centering
  \resizebox{\textwidth}{!}{%
  \input{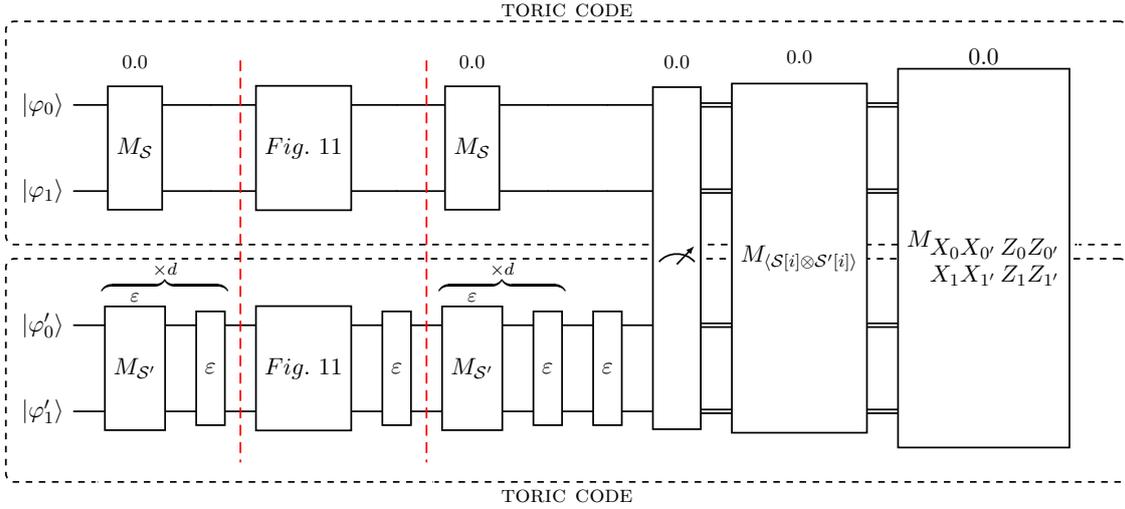}%

  }
  \caption{Folded \(S\) gate simulation protocol.}
  \label{fig:transS}
\end{figure*}

\begin{figure}
  \centering
  \resizebox{\linewidth}{!}{%
    \includegraphics{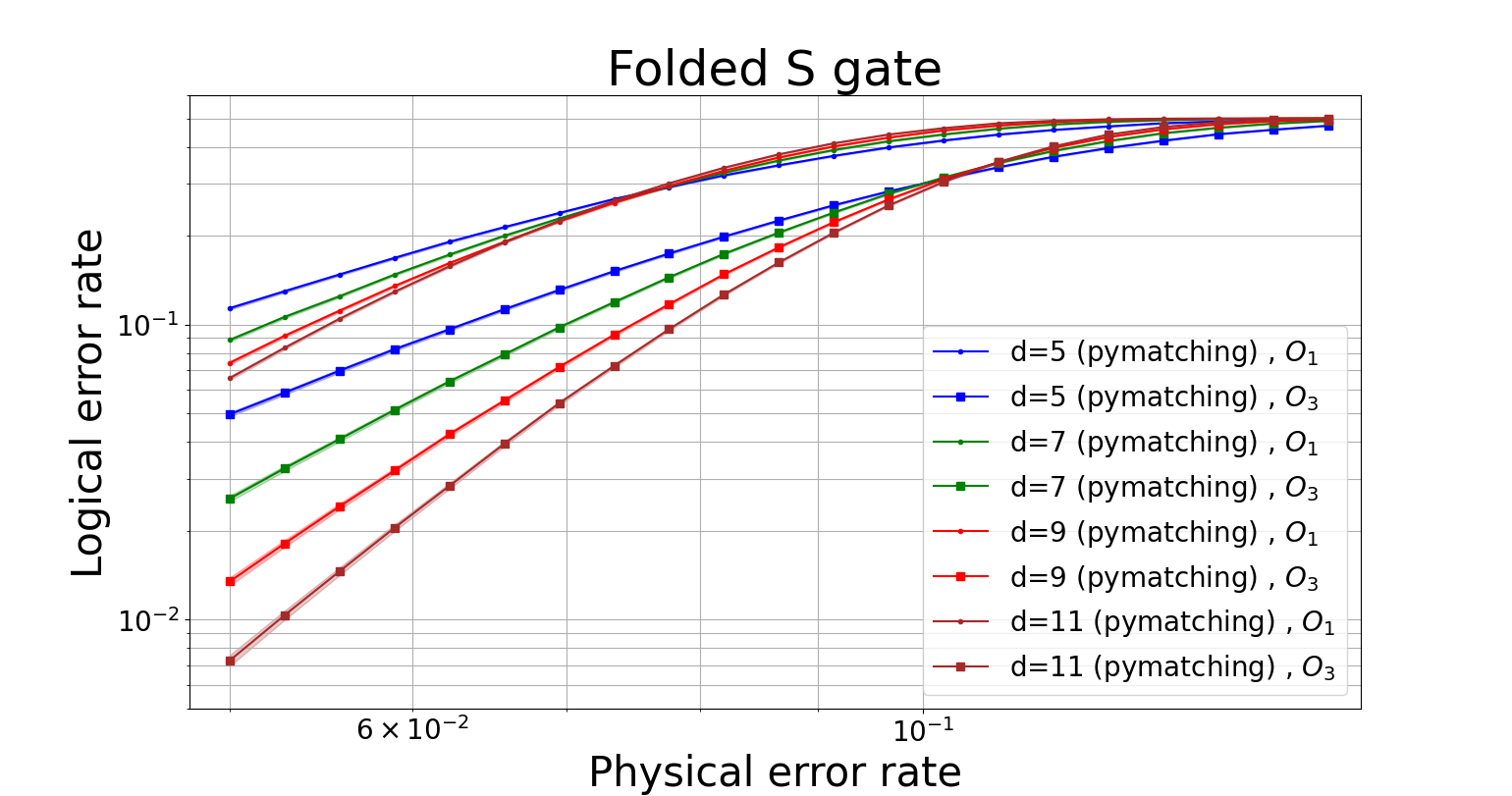}
  }
  \caption{Logical error rate estimation of the folded \(S\) gate using \texttt{pymatching}. The two tracked observable are $Y_0Y^{\prime}_0$ and $Z_0Z^{\prime}_0$ becoming $X_0X^{\prime}_0$ and $Z_0Z^{\prime}_0$ throughout computation.}
  \label{fig:TransSresults}
\end{figure}

\begin{figure}
  \centering
  \resizebox{\linewidth}{!}{%
    \includegraphics{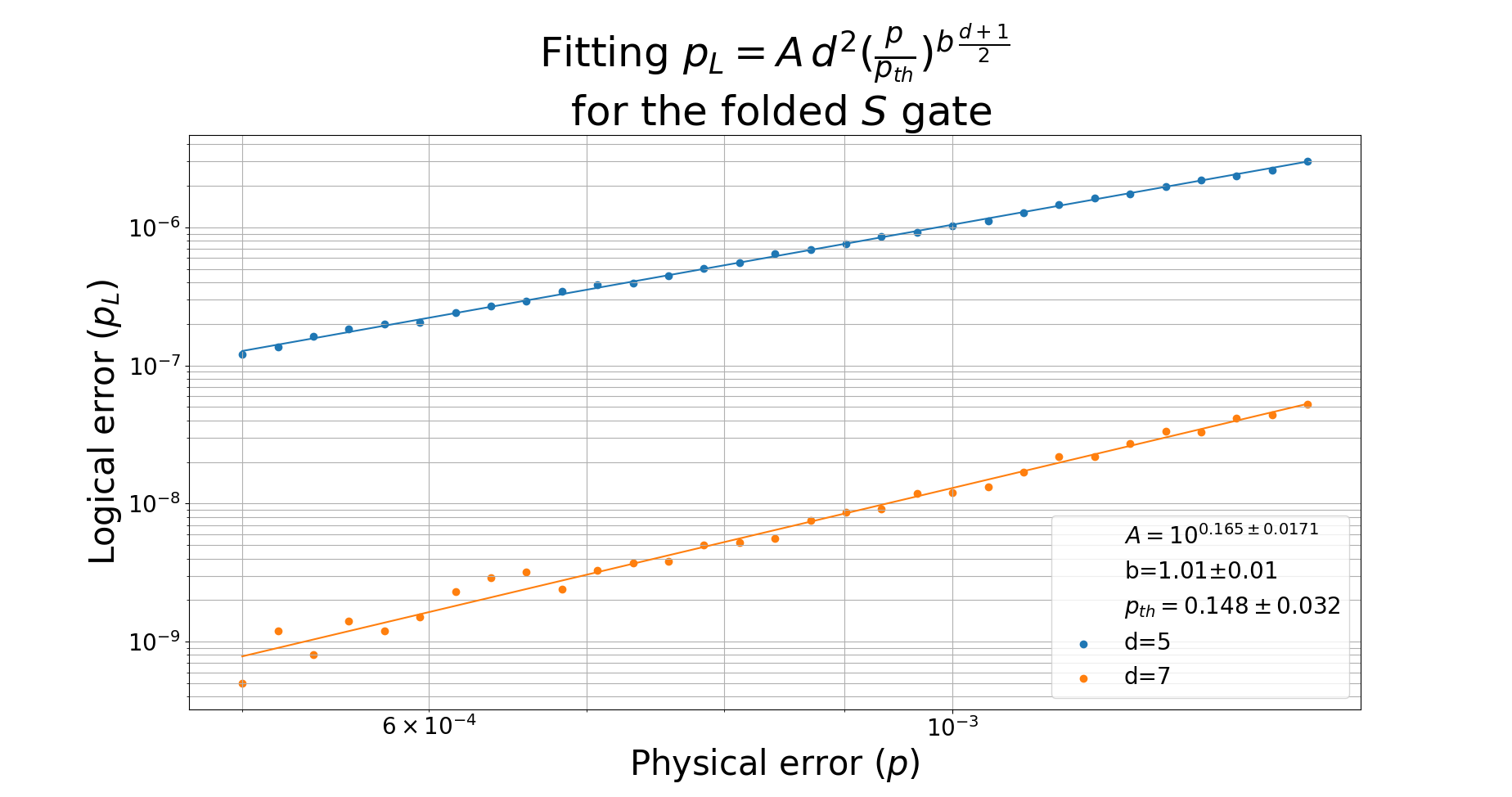}
  }
  \caption{Folded \(S\) marginal logical error rate fitting for low physical error rate.}
  \label{fig:fitTransS}
\end{figure}

The folded \(S\) logical error rates are estimated through the protocol described on Figure~\ref{fig:transS}. We start with one physical noisy toric code and one ideal noiseless toric code. Qubit states are entangled by pairs such that one qubit of each pair lies in the noisy code and the other in the noiseless code: $\left(\ket{\varphi_{1}\varphi_{1}^{\prime}} = \ket{\varphi_{2}\varphi_{2}^{\prime}} = \frac{\ket{00} + \ket{11}}{\sqrt{2}}\right)$. The final step's reconstructed logical measurements confronts respectively initial $Y_0Y^{\prime}_0$, $Y_1Y^{\prime}_1$, $Z_0Z_0^{\prime}$ and $Z_1Z^{\prime}_1$ with ending $X_0X^{\prime}_0$, $X_1X^{\prime}_1$, $Z_0Z_0^{\prime}$ and $Z_1Z^{\prime}_1$ to form the four observables.
We obtain a threshold of 7~\% \fig{TransSresults}.
The fit along model function \(f_{A, b, p_{th}}\) for the folded \(S\) gate protocol at low physical error rate features an effective distance reduction \(b\) close to 1 \fig{fitTransS}, indicating that code distance wasn't impacted during the procedure. The effective threshold (15\%) is still highly optimistic compared to the results shown on Figure~\ref{fig:TransSresults}.

\begin{figure}
  \centering
  \resizebox{\linewidth}{!}{%
    \includegraphics{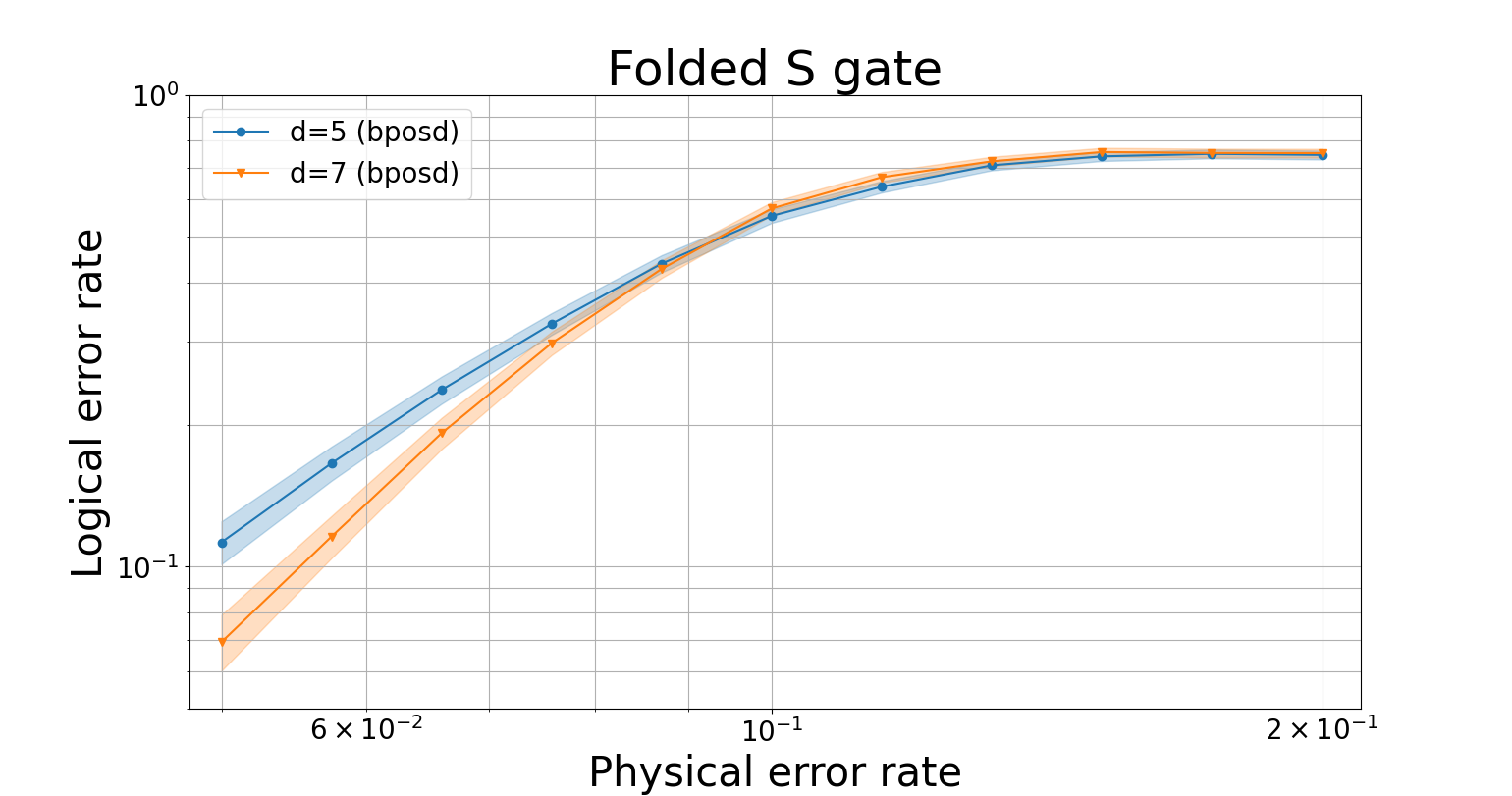}
  }
  \caption{Logical error rate estimation of the folded \(S\) gate using \texttt{BPOSD}. The two tracked observable are $Y_0Y^{\prime}_0$ and $Z_0Z^{\prime}_0$ becoming $X_0X^{\prime}_0$ and $Z_0Z^{\prime}_0$ throughout computation. A logical error is counted when either or both observables are flipped.}
  \label{fig:TransSresults2}
\end{figure}

\clearpage
\section{Conclusion}

We showed a complete set of constant-depth Clifford gates for the toric code, under the assumption that the cost of ancilla preparation can be amortized across the computation. We used a custom decoder that handles syndrome decoding multigraph by splitting the decoding along the hyperedges introduced by the logical gates. It was used to perform a threshold estimation of each gate, but would have to be modified to handle larger circuits with multiple logical gates chained. The numerical simulations for each gate exhibit memory-like behavior with the presence of a relatively high threshold.

Overall, all the protocols we fitted at low physical error rate either hinted a preserved distance with respect to the underlying toric code or clearly showed the need of a more resource-consuming investigation. The effective threshold derived from those fits are all highly optimistic, but we can postulate from those that the bottleneck operation among the logical gates we list is the partial measurement operation. Most difficulties we encountered while fitting these protocols can probably be explained by the hybrid error model we chose (circuit-level for logical gates and phenomenological for stabilizer measurements). One might want to refine the error model before pushing numerical studies further. Indeed, the high required connectivity will restrict the architectures that could implement these procedures which \textit{de facto} bring their own error model with them.

In this work, we focused on the toric code as the prime example of block codes encoding more than one qubit. Analogous procedures to the ones we studied might possibly be derived for some generalizations of the toric code, such as the hyperbolic surface codes. Indeed, such code are generally thought of as lattice on \(k\)-torus, with \(k\) larger than 1.

\section*{Acknowledgments}
\label{sec:ack}

All the quantum circuits from this paper were drawn with the \LaTeX\ package \texttt{quantikz}~\cite{kayTutorialQuantikzPackage2023}.
We acknowledge support from the Plan France 2030 through the project NISQ2LSQ ANR-22-PETQ-0006, HQI ANR-22-PNCQ-0002, from Inria EQIP challenge and Inria EA QASAR.

\clearpage
\clearpage
\bibliographystyle{quantum}

\bibliography{main}{}

\clearpage
\appendix

\section{Error Correlations}
\label{app:corr}
In this section, we look at the correlation that arises between logical errors on protected qubits of a toric code. We take advantage of the possibility to define complex observables in the \texttt{stim} framework to estimate the rate at which correlated errors happen. To detect a correlated error that unrecoverably flips two observables $O_1$ and $O_2$, we need to define and observe a third observable $O_3$ whose support is the symmetric difference of the support of $O_1$ and $O_2$. An error unrecoverably flipping either $O_1$ or $O_2$ but not both is not a correlated logical error, and \emph{will} flip $O_3$ whereas an error unrecoverably flipping both $O_1$ and $O_2$ is a correlated logical error and will \emph{not} flip $O_3$. Hence, by summing over all logical errors flipping $O_1$ or $O_2$ and subtracting errors that flipped $O_3$, we end up counting twice each correlated error that happened on the logical observables. In other words, denoting by $E_i$ the event where the observable $O_i$ has been flipped :
\resizebox{\textwidth}{!}{$2 P\left(E_1 \cap E_2\right) = P\left(E_1 \cap E_2\right) + P\left(E_1 \cap \overline{E_2}\right) + \mathbf{P\left(E_1 \boldsymbol{\cap} E_2\right) + P\left(\overline{E_1} \boldsymbol{\cap} E_2\right)} - P\left(E_1 \cap \overline{E_2}\right) - P\left(\overline{E_1} \cap E_2\right) =P\left(E_1\right) + \mathbf{P\left(E_2\right)} - P\left(E_3\right)$}

These three probabilities are precisely what is estimated by sampling in an error rate simulation experiment.

We looked at the correlations in different scenarios: \((Z_1,Z_2)\) in a memory experiment; \((Y_1, Y_2)\), \((Z_1, Z_2)\) and \((Z_1, Y_2)\) in a folded \(S\) gate experiment; \((Z_1,Z_2)\) for the regular and instantaneous Dehn twists.
Mostly we observe little correlations as soon as we depart from the threshold but we do observe a bump around the threshold value.
The Dehn twists are a little bit different: we see higher correlations and a slower decrease away from threshold, which could be understood by the fact that the logical operation performed is supposed to propagate logical errors where the others are not. 

These observations are reassuring in the case of the folded \(S\) gate as even with intra-block two-qubit gates we don't see more correlated errors introduced than a memory experiment.

\begin{figure}
  \centering
  \resizebox{\linewidth}{!}{%
    \includegraphics{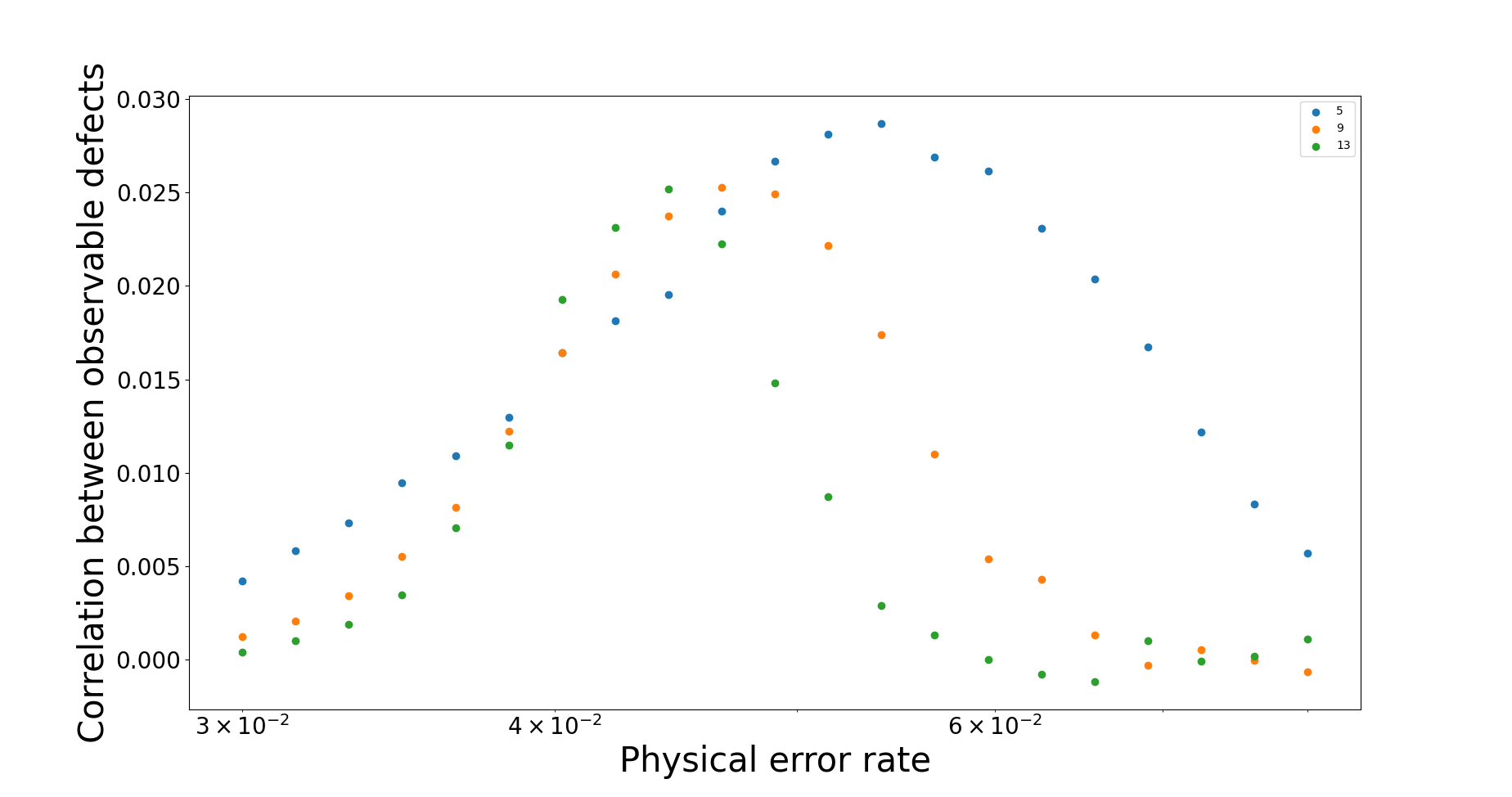}
  }
  \caption{Correlation between $X_1$ logical error rate and $X_2$ logical error rate of toric code in a memory experiment.}
  \label{fig:corrmemory}
\end{figure}

\begin{figure}
  \centering
  
    \includegraphics[width=.5\linewidth]{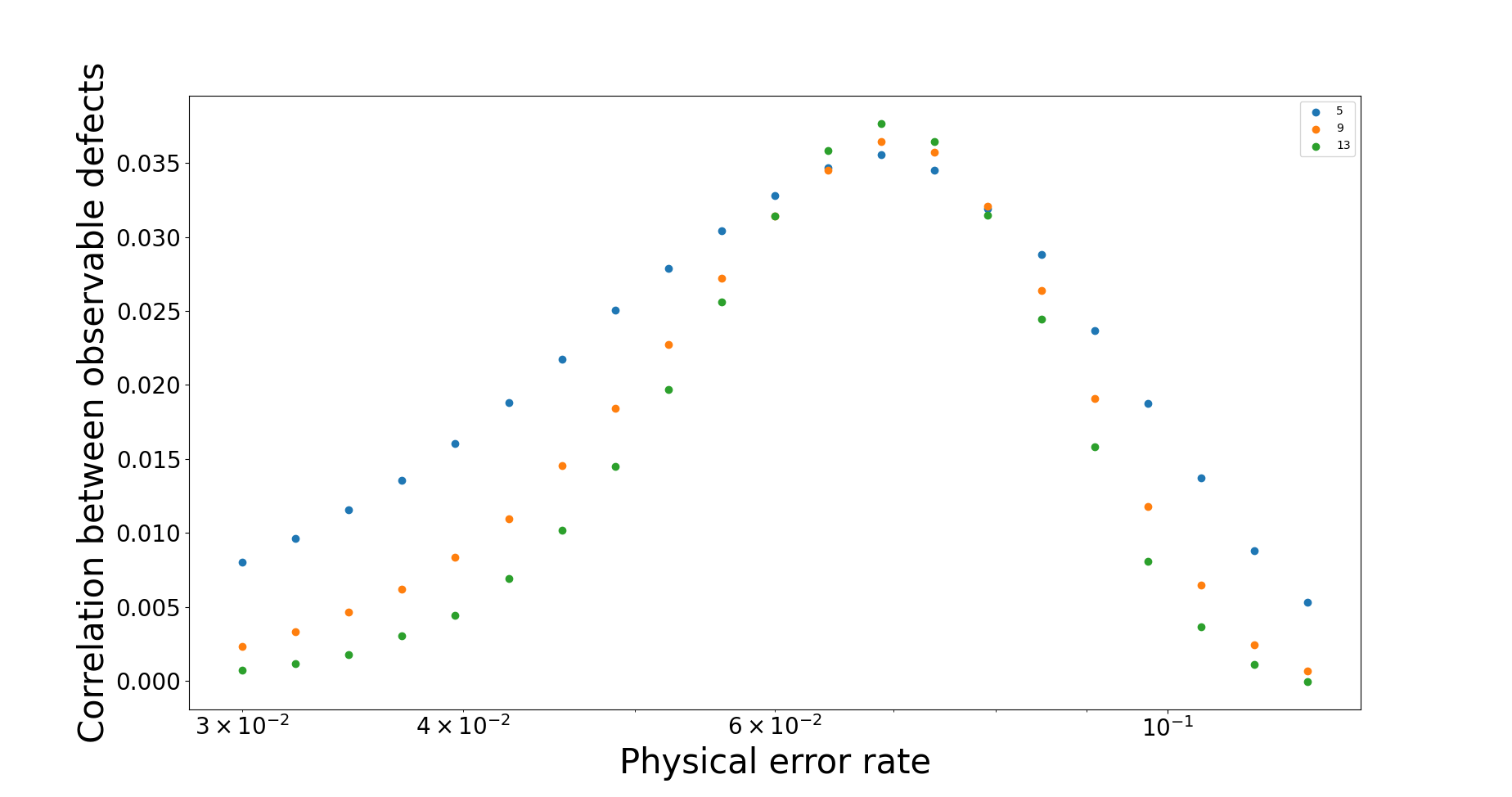}\hfill 
  \includegraphics[width=.5\linewidth]{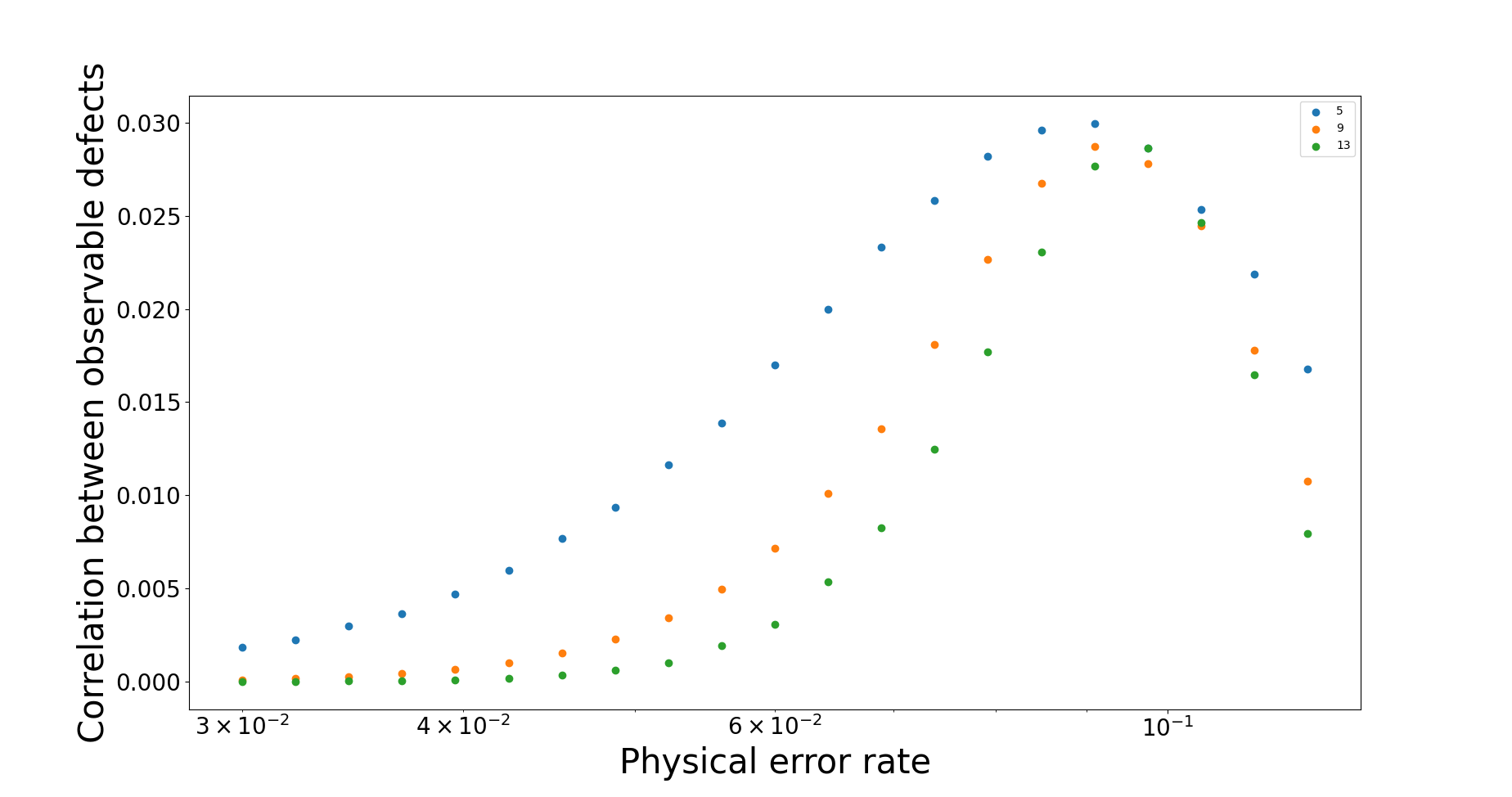}
  \hfill
\includegraphics[width=.5\linewidth]{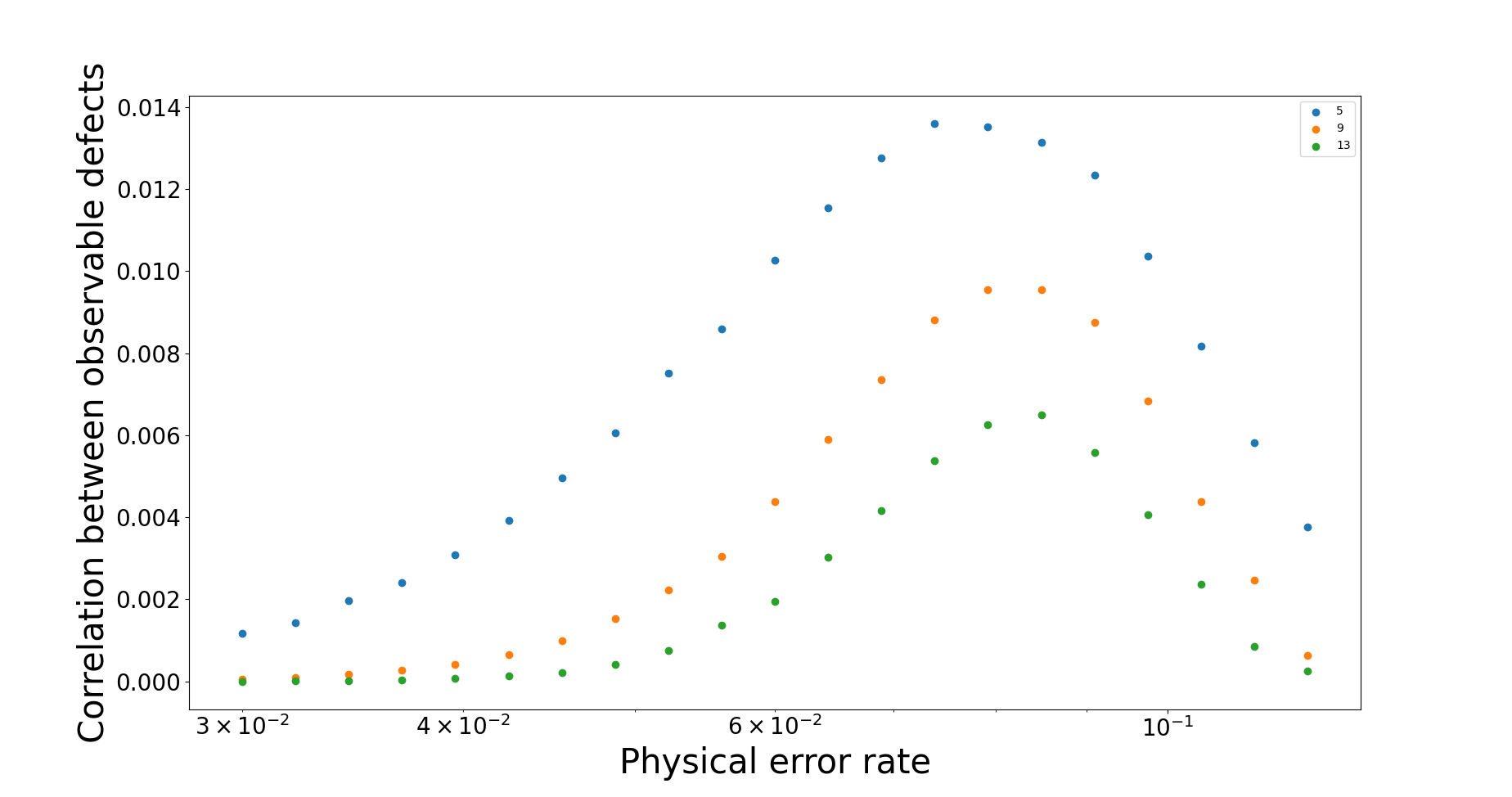}

  \caption{Correlation between $Y_1$ and $Y_2$ (top left), $Z_1$ and $Z_2$ (top right), $Z_1$ and $Y_2$ (bottom) of a folded \(S\) gate.}
  \label{fig:corrtransS}
\end{figure}

\begin{figure}
  \centering
    \includegraphics[width=.5\linewidth]{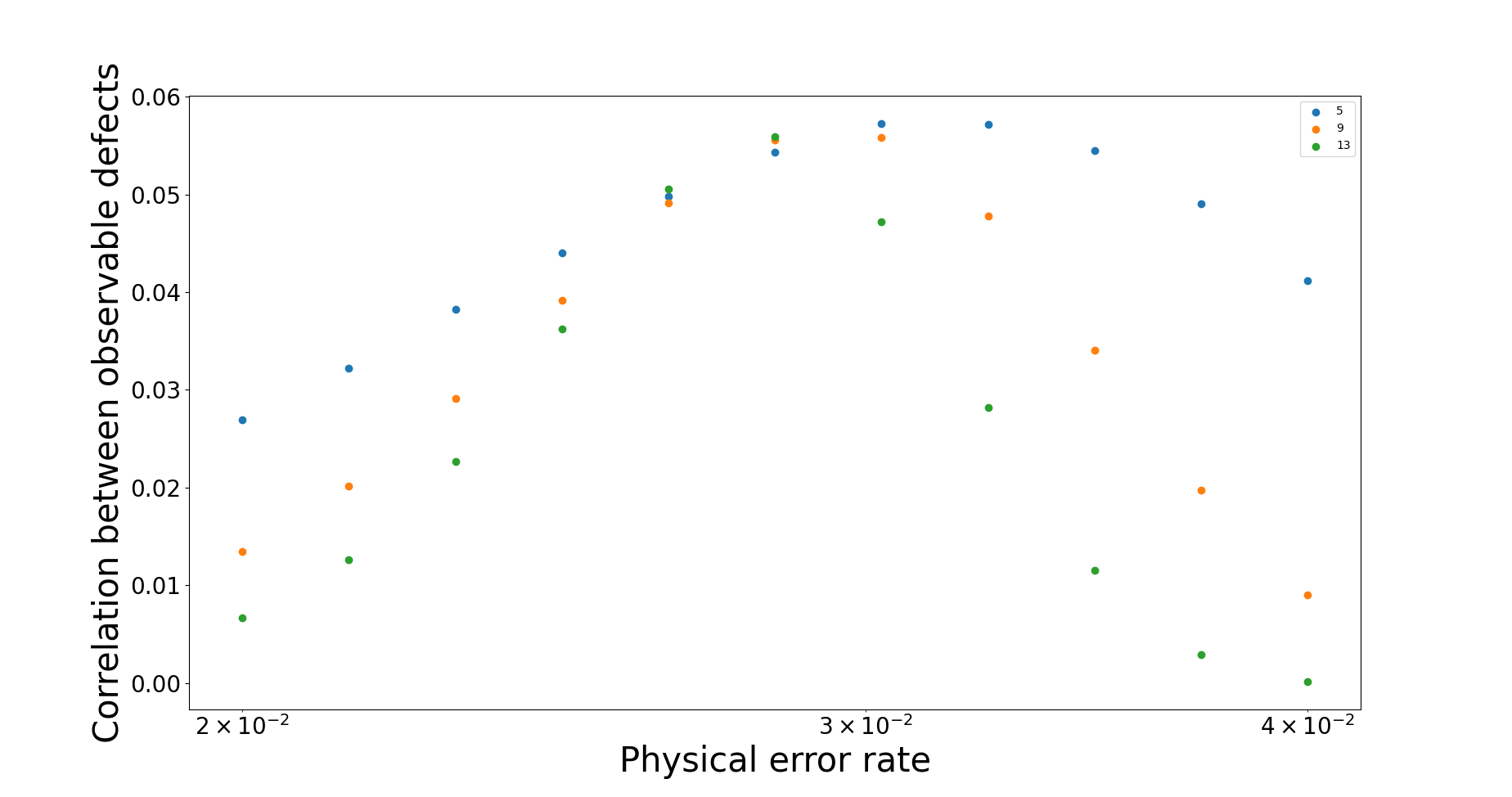}\hfill
  \includegraphics[width=.5\linewidth]{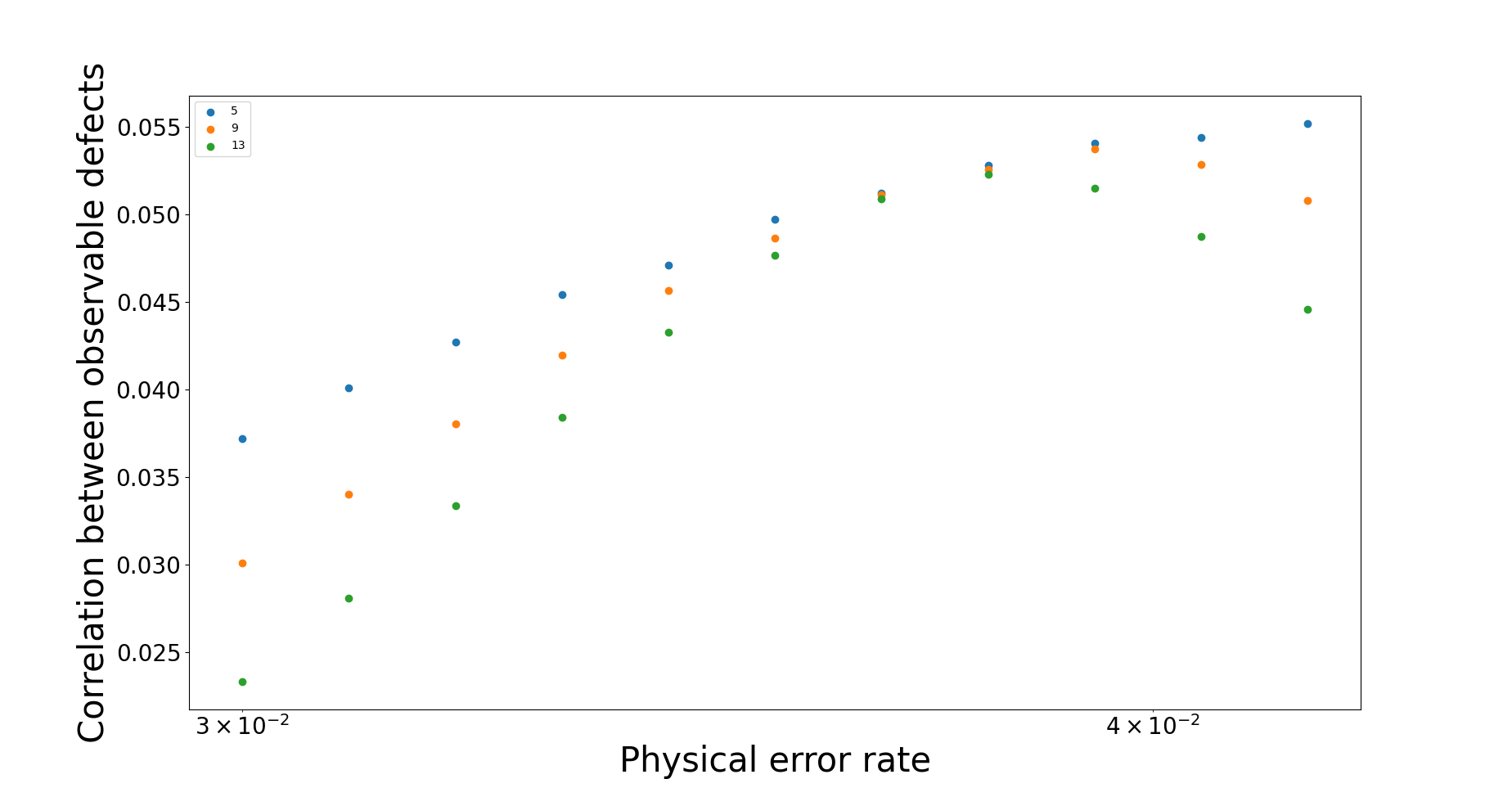}
  \caption{Correlation between $Z_1$ logical error rate and $Z_2$ logical error rate of a standard Dehn twist (left) and instantaneous Dehn twist (right).}
  \label{fig:corrIDT}
\end{figure}

\clearpage

\end{document}